\documentclass[12pt]{article}

\newcommand{\be}{\begin{equation}}
\newcommand{\ee}{\end{equation}}
\newcommand{\bea}{\begin{eqnarray}}
\newcommand{\eea}{\end{eqnarray}}
\newcommand{\beaa}{\begin{eqnarray*}}
\newcommand{\eeaa}{\end{eqnarray*}}

\newcommand{\g}{{\bf g}}
\newcommand{\h}{{\bf h}}
\newcommand{\kk}{{\bf k}}
\newcommand{\qa}{\mathbf{a}}
\newcommand{\qe}{\mathbf{e}}
\newcommand{\qi}{\mathbf{i}}
\newcommand{\qj}{\mathbf{j}}
\newcommand{\qk}{\mathbf{k}}
\newcommand{\ma}{\mathbb{A}}
\newcommand{\me}{\mathbb{E}}
\newcommand{\mi}{\mathbb{I}}
\newcommand{\mj}{\mathbb{J}}
\newcommand{\mk}{\mathbb{K}}

\advance\textheight by \topskip
\usepackage{amssymb}
\usepackage{pstricks}
\newcommand{\psl}{\psline[linewidth=.01]}
\newcommand{\one}{\begin{pspicture}(0,.1)(.5,.6)
\psline(0,.25)(.5,.25)
\end{pspicture} }
\newcommand{\ons}{\begin{pspicture}(0,.1)(.5,.6)
\psline(0,.25)(.5,.25)
\pscircle[fillstyle=solid](.25,.25){.08}
\end{pspicture} }
\newcommand{\onsc}{\begin{pspicture}(0,.1)(.5,.6)
\psline(0,.25)(.5,.25)
\pscircle[fillstyle=solid,fillcolor=black](.25,.25){.08}
\end{pspicture} }
\newcommand{\onjr}{\begin{pspicture}(0,.1)(.5,.6)
\psline(0,.25)(.5,.25)
\psline(.3,.25)(.2,.35)
\psline(.3,.25)(.2,.15)
\end{pspicture} }
\newcommand{\onjl}{\begin{pspicture}(0,.1)(.5,.6)
\psline(0,.25)(.5,.25)
\psline(.2,.25)(.3,.35)
\psline(.2,.25)(.3,.15)
\end{pspicture} }
\newcommand{\onul}{\begin{pspicture}(0,.1)(.25,.6)
\psline(0,0)(.08,.07)
\psline(.08,.07)(.13,.14)
\psline(.13,.14)(.15,.21)
\psline(.15,.21)(.15,.29)
\psline(.15,.29)(.13,.36)
\psline(.13,.36)(.08,.43)
\psline(.08,.43)(0,.5)
\end{pspicture} }
\newcommand{\onur}{\begin{pspicture}(.25,.1)(.5,.6)
\psline(.5,0)(.42,.07)
\psline(.42,.07)(.37,.14)
\psline(.37,.14)(.35,.21)
\psline(.35,.21)(.35,.29)
\psline(.35,.29)(.37,.36)
\psline(.37,.36)(.42,.43)
\psline(.42,.43)(.5,.5)
\end{pspicture} }
\newcommand{\onuls}{\begin{pspicture}(0,.1)(.25,.6)
\psline(0,0)(.08,.07)
\psline(.08,.07)(.13,.14)
\psline(.13,.14)(.15,.21)
\psline(.15,.21)(.15,.29)
\psline(.15,.29)(.13,.36)
\psline(.13,.36)(.08,.43)
\psline(.08,.43)(0,.5)
\pscircle[fillstyle=solid](.15,.25){.08}
\end{pspicture} }

\newcommand{\onjdl}{\begin{pspicture}(0,.1)(.25,.6)
\psline(0,0)(.08,.07)
\psline(.08,.07)(.13,.14)
\psline(.13,.14)(.15,.21)
\psline(.15,.21)(.15,.29)
\psline(.15,.29)(.13,.36)
\psline(.13,.36)(.08,.43)
\psline(.08,.43)(0,.5)
\psline(.15,.2)(.05,.3)
\psline(.15,.2)(.25,.3)
\end{pspicture} }
\newcommand{\onjur}{\begin{pspicture}(.25,.1)(.5,.6)
\psline(.5,0)(.42,.07)
\psline(.42,.07)(.37,.14)
\psline(.37,.14)(.35,.21)
\psline(.35,.21)(.35,.29)
\psline(.35,.29)(.37,.36)
\psline(.37,.36)(.42,.43)
\psline(.42,.43)(.5,.5)
\psline(.35,.3)(.25,.2)
\psline(.35,.3)(.45,.2)
\end{pspicture} }
\newcommand{\tws}{\begin{pspicture}(0,.1)(.5,.6)
\psline(0,0)(.5,0)
\psline(0,.5)(.5,.5)
\end{pspicture} }
\newcommand{\twc}{\begin{pspicture}(0,.1)(.5,.6)
\psline(0,0)(.5,.5)
\psline(0,.5)(.5,0)
\end{pspicture} }
\newcommand{\twu}{\onul \onur}
\newcommand{\twj}{\onjdl \onjur}
\newcommand{\twssu}{\begin{pspicture}(0,.1)(.5,.6)
\psline(0,0)(.5,0)
\psline(0,.5)(.5,.5)
\pscircle[fillstyle=solid](.25,.5){.08}
\end{pspicture} }
\newcommand{\twsscu}{\begin{pspicture}(0,.1)(.5,.6)
\psline(0,0)(.5,0)
\psline(0,.5)(.5,.5)
\pscircle[fillstyle=solid,fillcolor=black](.25,.5){.08}
\end{pspicture} }
\newcommand{\trs}{\begin{pspicture}(0,.1)(.5,.6)
\pscircle(.25,.25){.25}
\pscircle[fillstyle=solid](.423,.423){.08}
\end{pspicture} }
\newcommand{\ctradn}{\begin{pspicture}(0,.1)(.4,.5)
\psline(.07,0)(.4,.33)
\psline(0,.07)(.33,.4)
\psline(0,.33)(.33,0)
\psline(.07,.4)(.4,.07)
\end{pspicture} }
\newcommand{\vsu}{\begin{pspicture}(0,0)(.05,.35)
\end{pspicture}}
\newcommand{\vs}{\begin{pspicture}(0,0)(.05,.55)
\end{pspicture}}
\newcommand{\vse}{\begin{pspicture}(0,0)(.05,.75)
\end{pspicture}}
\makeatletter

\@addtoreset{equation}{section}

\def\section{\@startsection {section}{1}{\z@}{-3.5ex plus -1ex minus
 -.2ex}{2.3ex plus .2ex}{\large\bf\centering}}
\def\subsection{\@startsection{subsection}{2}{\z@}{-3.25ex plus%
 -1ex minus -.2ex}{1.5ex plus .2ex}{\bf}}
\def\subsubsection{\@startsection{subsubsection}{3}{\z@}{-3.25ex plus%
 -1ex minus -.2ex}{1.5ex plus .2ex}{\sl}}
\makeatother

\begin{document}

\baselineskip 17pt
\parindent 10pt
\parskip 9pt

\begin{titlepage}
\begin{flushright}
hep-th/0104212\\ April 2001\\[3mm]
\end{flushright}
\vspace{2cm}
\begin{center}
{\Large {\bf Boundary scattering, symmetric spaces and the\\
principal chiral model on the half-line}}\\ \vspace{1cm} {\large
N.J. MacKay\footnote{\tt nm15@york.ac.uk}, B. J.
Short\footnote{\tt bjs108@york.ac.uk}}
\\
\vspace{3mm} {\em Department of Mathematics,\\ University of York,
\\York YO10 5DD, U.K.}
\end{center}

\vspace{1.5cm}
\begin{abstract}
\noindent We investigate integrable boundary conditions (BCs) for
the principal chiral model on the half-line, and rational
solutions of the boundary Yang-Baxter equation (BYBE). In each
case we find a connection with (type I, Riemannian, globally)
symmetric spaces $G/H$: there is a class of integrable BCs in
which the boundary field is restricted to lie in a coset of $H$;
these BCs are parametrized by ${G/ H}\!\times{G/ H}$; there are
rational solutions of the BYBE in the defining representations of
all classical $G$ parametrized by $G/H$; and using these we
propose boundary $S$-matrices for the principal chiral model,
parametrized by ${G/ H}\!\times{G/ H}$, which correspond to our
boundary conditions.
\end{abstract}

\end{titlepage}

\section{Introduction}

The bulk principal chiral model (PCM) -- that is, the
$1+1$-dimensional, $G\!\times G$-invariant nonlinear sigma model
with target space a compact Lie group $G$ -- is known to have a
massive spectrum of particles in multiplets which are (sometimes
reducible) representations of $G\!\times G$. These are irreducible
representations, however, of the Yangian algebra of non-local
conserved charges, and the multiplets are also distinguished by a
set of local, commuting conserved charges with spins equal to the
exponents of $G$ modulo its Coxeter number. The corresponding bulk
scattering (`$S$-') matrices are constructed from $G$-invariant
(rational) solutions of the Yang-Baxter equation (YBE), whose
poles determine the couplings between the multiplets.


In this paper we investigate the model on the half-line -- that
is, with a boundary. Any proposed boundary $S$-matrices must
satisfy the boundary Yang-Baxter equation (BYBE), for which only a
limited range of solutions is known
(\cite{chere83,macka95,abad95,gand99,dev92,lima99} is a
selection). For constant solutions of the BYBE ({\em i.e.\
}without dependence on a spectral parameter or rapidity) there is
a well-established connection with (quantum) symmetric spaces
\cite{qss}.

We shall find a class of BYBE solutions corresponding to, and
parametrized by, the symmetric spaces $G/H$. These solutions
utilize for the bulk S-matrix the rational\footnote{before the
inclusion of scalar prefactors}, $G$-invariant solution of the
(usual, bulk) YBE in the defining ($N$-dimensional vector)
representation of a classical $G$, and are themselves rational and
$N$-dimensional, describing the scattering of the bulk vector
particle off the boundary ground state. We make the ansatz that
they are constant or linear in rapidity, and thus have at most two
channels. The underlying algebraic structures are the twisted
Yangians \cite{molev96}, though the relationship remains to be
explored.

However, we begin by investigating how our solutions might arise
as boundary $S$-matrices, by discussing the principal chiral field
on the half-line and boundary conditions which preserve its
integrability (see also \cite{corr97}, and \cite{corr96} for a
more general discussion of boundary integrability). We shall find
two classes of BCs which are associated with the $G/H$. In the
first, `chiral' class, the field takes its values at the boundary
in a coset of $H$, and the space of such cosets is (up to a
discrete ambiguity) ${G\over H}\!\times\!{G\over H}$.
Correspondingly, we use our BYBE solutions to construct boundary
$S$-matrices, parametrized by ${G\over H}\!\times\!{G\over H}$,
which preserve the same remnant of the $G\!\times G$ symmetry as
the integrable boundary conditions. In the second, `non-chiral'
class, for which we do not generally have corresponding boundary
$S$-matrices, the boundary field lies in a translate of the Cartan
immersion of $G/H$ in $G$. To summarize: a connection between
boundary integrability and symmetric spaces emerges naturally in
two very different ways: by seeking classically integrable
boundary conditions, and by solving the BYBE.

The plan of the paper is as follows. In section two, building
naturally on the results of \cite{evans97,evans99} for the bulk
PCM, we discuss boundary conditions which lead naturally to
conservation of local charges. As mentioned, there are two classes
of BC, which we call `chiral' and `non-chiral'. In section three
we find minimal boundary $S$-matrices, by making ans\"atze for the
BYBE solutions and applying the conditions of crossing-unitarity,
hermitian analyticity and $R$-matrix unitarity, and explain how
these are related to symmetric spaces. This section is necessarily
rather long and involved, and many of the details appear in
appendices. From these, in section four, we construct boundary
$S$-matrices for the PCM, and find that these correspond naturally
to the chiral BCs. The key statements of our results for the
boundary $S$-matrices can be found in section~\ref{subsec:bsm}
(for the minimal case, without physical strip poles) and
section~\ref{subsec:tbsm} (for the full PCM $S$-matrices). This
paper supersedes the preliminary work of \cite{macka99}.

\section{The principal chiral model on the half-line}
\label{sec:pcmohl}

\subsection{The principal chiral model on the full line}

We first describe the model on the full line, without boundary.
This subsection is largely drawn from \cite{evans99}, and full
details may be found there.

The principal chiral model may be defined by the lagrangian
\be\label{pcmlagr} {\cal L} = {1\over 2}{\rm Tr}\left(
\partial_\mu g^{-1}
 \partial^\mu g\right) \, ,
\ee where the field $g(x^\mu)$ takes values in a compact Lie group
$G$. (We could also include an overall, coupling constant, but
this may be absorbed into $\hbar$, and will not be important for
our purposes.) It has a global $G_L\times G_R$ symmetry $g\mapsto
g_L g g_R^{-1}$ associated with conserved currents
\be\label{lrcurr} j(x,t)_\mu^L=\partial_{\mu} g \,g^{-1} , \qquad
j(x,t)_\mu^R = - g^{-1}\partial_{\mu} g \ee which take values in
the Lie algebra $\g$ of $G$: that is, $j=j^a t^a$ (for $j^L$ or
$j^R$: henceforth we drop this superscript) where $t^a$ are the
generators of $\g$, and (with $G$ compact)
Tr$(t^at^b)=-\delta^{ab}$. The equations of motion are
\be\label{cons}
\partial^{\mu}j_{\mu}(x,t)=0 \, , \hspace{0.4in}
\partial_{\mu} j_{\nu} - \partial_{\nu} j_{\mu} - [j_{\mu},j_{\nu}] = 0 \,,
\ee which may be combined as \be\label{lce}
\partial_- j_+ = - \partial_+ j_- = - {1\over 2} [j_+,j_-]
\ee in light-cone coordinates $x^\pm = {1\over 2} (t \pm x)$ (and
thus $\partial_\pm = \partial_0 \pm \partial_1$).

In addition to the usual spatial parity $P:x\mapsto -x$, the PCM
lagrangian has further involutive discrete symmetries. The first,
which we call $G$-parity and which exchanges $L\leftrightarrow R$,
is \be\label{parity} \pi \, : \; g\mapsto g^{-1} \;\; \Rightarrow
 \;\; j^L \leftrightarrow j^R \,.\ee
 (In the usual QCD effective model, `parity' is the combination $P\pi$.)
Then there is $g\mapsto\alpha(g)$ where $\alpha$ is any involutive
automorphism, though only for outer automorphisms may this have a
non-trivial effect on the invariant tensors and local charges
which we shall consider shortly.

The canonical Poisson brackets for the model are
\begin{eqnarray}
\left\{ j_0^a (x), j_0^b (y) \right\} & = & f^{abc} \, j_0^c (x)
\, \delta(x{-}y) \nonumber \\ \left\{ j_0^a (x), j_1^b (y)
\right\} & = & f^{abc} \, j_1^c (x) \, \delta(x{-}y) + \,
\delta^{ab} \, \delta'(x{-}y) \label{PBs}\\ \left\{ j_1^a (x),
j_1^b (y) \right\} & = & 0 \nonumber
\end{eqnarray}
at equal time. These expressions hold for {\em either\/} of the
currents $j^L$ or $j^R$ separately, while the algebra of $j^L$
with $j^R$ (which we shall not need here) involves only
$\delta'(x{-}y)$ terms in the brackets of space- with
time-components.

This model has two distinct sets of conserved charges, and the two
sets commute. The first is the extension of the $G_L\times G_R$
charges to the larger algebra of non-local, Yangian charges
\cite{lusch78,berna91} $Y(\g_L)\times Y(\g_R)$; we shall not
discuss these here. There is also\footnote{In this paper we
restrict to the classical $\g$, although they also exist for
exceptional $\g$ \cite{evans01}.} an infinite set of local,
commuting charges with spins $s$ equal to the exponents of $\g$
modulo its Coxeter number, \be \label{locch} q_{\pm s} \,=\,
\int_{-\infty}^\infty k_{a_1a_2 \ldots a_n} \, j_\pm^{a_1}(x)
j_\pm^{a_2} (x) \ldots j_\pm^{a_n}(x) \,dx \ee (where $n=s+1$);
here, unlike for the Yangian, $j^L$ and $j^R$ give the same
charges (up to a change of sign), of which there is therefore only
one set. The primitive invariant tensors $k$ have to be very
carefully chosen to ensure the charges commute -- for the full
story see \cite{evans99}. Such a set appears to be precisely what
is needed for quantum integrability, where it leads to the
beautiful structure of masses and interactions described in
\cite{dorey91}.

\subsection{Boundary conditions for the model on the half-line}

Varying the bulk action on the half-line $-\infty<x\leq 0$ imposes
the additional boundary equation $$ {\rm Tr}(g^{-1} \partial_1
g.g^{-1}\delta g)=0 \qquad {\rm at} \;\;\; x=0\,,$$ where the
variation is over all $\delta g$ such that $g^{-1}\delta g\in\g$.
Clearly the Neumann condition $\partial_1 g|_0=0$ solves this, as
does the Dirichlet condition $\delta g|_0=0$, or
$\partial_0g|_0=0$. But we can also impose mixed conditions, in
any way such that $$ (g^{-1}\partial_1 g)^a\,(g^{-1}\partial_0
g)^a=0$$ (with the usual summation convention).

We begin by considering some simple mixed boundary conditions
written in terms of the currents $j$. A little later we shall
generalize these, and write them in terms of the fields $g$. We
take as a BC on the currents \be\label{bc1} j_+^a(0) =
R^{ab}j_-^b(0)\,,\ee with each $j$ chosen independently to be
either L or R; we refer to the four possibilities as LL, RR, LR
and RL. The boundary equation of motion then requires that
$R^{ab}$ be an orthogonal matrix. We would also like consistency
with the Poisson brackets: if we extend the currents' domains to
$x>0$ by requiring $ j_+^a(x) = R^{ab}j_-^b(-x)$, then this
further requires that $R$ be symmetric, and give an (involutive;
$\alpha^2=1$) automorphism $\alpha$ of $\g$ via $$\alpha :
t^a\mapsto R^{ab}t^b\,.$$ Together these imply that $R$ is
diagonalizable with eigenvalues $\pm 1$, so that we may write $$
\g=\h\oplus\kk\;,$$ where $\h$ and $\kk$ are the $+1$ and $-1$
eigenspaces respectively. The $\h$ indices then correspond to
Neumann directions $j_+^a=j_-^a \;\Rightarrow\; j_1^a=0$, the
$\kk$ indices to Dirichlet directions $j_+^a=-j_-^a
\;\Rightarrow\; j_0^a=0$ (all at $x=0$). Further, $R$'s being an
automorphism implies that $$ [\h,\h]\subset\h\;,\qquad
[\h,\kk]\subset \kk \;,\qquad [\kk,\kk]\subset \h\,,$$ precisely
the properties required of a symmetric space $G/H$ \cite{helga78},
where $H$ is the subgroup generated by $\h$ and invariant under
the involution $\alpha$.

For integrability we require a great deal of $R$. As we have said,
there are two infinite sets of charges. The Yangian charges appear
no longer to be conserved on the half-line, even with pure Neumann
BCs \cite{mour95} (naively, at least, it seems that there are
remnants only, as we shall see). However, we believe they are not
essential for integrability, because precisely half of the local
charges remain conserved, with either $q_s+q_{-s}$ or $q_s-q_{-s}$
surviving. Our conjecture is that these are enough to guarantee
the properties of quantum integrability, such as factorizability
of the $S$-matrix.

The first charge is energy, and its conservation on the half-line
is just the equation of motion, requiring $R$ to be orthogonal.
For the higher charges we take \beaa q_{|s|} & = & q_s \pm q_{-s}
\\ & = &  \int_{-\infty}^0 k_{a_1 a_2...a_n}\left\{
j_+^{a_1}...j_+^{a_n} \pm j_-^{a_1}...j_-^{a_n}\right\}\,dx \,.
\eeaa Then \beaa {d\over dt} q_{|s|} & = &\int_{-\infty}^0 k_{a_1
a_2...a_n}\left\{
(\partial_-+\partial_1)\left(j_+^{a_1}...j_+^{a_n}\right) \pm
(\partial_+-\partial_1) \left(
j_-^{a_1}...j_-^{a_n}\right)\right\}\,dx \\ & = & k_{a_1
a_2...a_n}\left\{ j_+^{a_1}...j_+^{a_n} \mp
j_-^{a_1}...j_-^{a_n}\right\} \mid_{x=0} \\ & = & \left(k_{b_1
b_2...b_n}R^{a_1 b_1}...R^{a_n b_n}\mp k_{a_1...a_n}\right)
j_-^{a_1}...j_-^{a_n})\mid_{x=0}\,. \eeaa That this is zero for
one choice of sign follows from the result \cite{evans00} that,
for every $R$ and $k$, \be\label{kinv} k_{b_1 b_2...b_n}R^{a_1
b_1}...R^{a_n b_n}= \epsilon k_{a_1...a_n}\,,\ee where
$\epsilon=\pm 1$. (This is obvious, with $\epsilon=1$, when
$\alpha$ is an inner automorphism, but not at all obvious for
outer automorphisms.) So, if we now regard $\alpha$ as acting on
the currents, $\alpha(j_\pm^a)=R^{ab}j_\pm^b$, we see that
\be\label{lch} q_{|s|}=\int_{-\infty}^0 k_{a_1 a_2...a_n}\left\{
j_+^{a_1}...j_+^{a_n} +
\alpha(j_-^{a_1})...\alpha(j_-^{a_n})\right\}\,dx\ee is the charge
which remains conserved in the presence of the boundary. For LL
and RR conditions its density is the combination which is
invariant under the combined action $P\alpha$ of spatial parity
$P$ (which exchanges $j_+\leftrightarrow j_-$) and $\alpha$, while
for LR and RL it is the combination invariant under these together
with $G$-parity, $P\alpha\pi$.

Further, these charges still commute. In the Poisson bracket of
the charges constructed from tensors $k^{(1)}$ and $k^{(2)}$, a
total derivative term which vanished in the bulk now gives an
additional contribution proportional to (all at $x=0$) \beaa &&
k^{(1)}_{ca_1...a_s}
k^{(2)}_{cb_1...b_r}\!\!\left(j_+^{a_1}...j_+^{a_s}j_+^{b_1}...j_+^{b_r}
\!-\epsilon^{(1)} \epsilon^{(2)}
j_-^{a_1}...j_-^{a_s}j_-^{b_1}...j_-^{b_r}\right) \\ &\!=
&\!k^{(1)}_{cd_1...d_s} k^{(2)}_{ce_1...e_r}\!\!\left(
R^{d_1a_1}...R^{d_sa_s}\,R^{e_1b_1}...R^{e_rb_r}\!-\epsilon^{(1)}
\epsilon^{(2)} \delta^{d_1a_1}...
\delta^{d_sa_s}\delta^{e_1b_1}...\delta^{e_rb_r}\right)
j_-^{a_1}...j_-^{a_s}j_-^{b_1}...j_-^{b_r} \\ &\!=
&\!k^{(1)}_{d_0d_1...d_s} k^{(2)}_{e_0e_1...e_r}\!\!\left(
R^{d_0a_0}R^{d_1a_1}...R^{e_0b_0}R^{e_1b_1}...\!-\epsilon^{(1)}
\epsilon^{(2)} \delta^{d_0a_0}\delta^{d_1a_1}...
\delta^{e_0b_0}\delta^{e_1b_1}... \right)\delta^{a_0b_0}
j_-^{a_1}...\;j_-^{b_r}
\\ &\!= &\!0 \eeaa by property (\ref{kinv}).
Finally, it is precisely the $q_{|s|}$ of (\ref{lch}) above that
still commute with the $G\!\times G$-generating charges
$Q=\int_{-\infty}^0 j_0^{L/R}\,dx$.

\subsubsection{General chiral BCs}
\label{ssubsec:phlcbc}

As we commented when first introducing the BC (\ref{bc1}), in
$j_+=\alpha(j_-)$ we may take each $j$ as either L or R. The LL
and RR conditions are then related:
$j_+^L=\alpha(j_-^L)\;\Rightarrow \;g j_+^R g^{-1} = \alpha(g
j_-^R g^{-1})$ (all at $x=0$). In fact the most general such (we
shall call it `chiral') BC is to take \be\label{bc2} g(0)\in k_L H
k_R^{-1}\,. \ee This is the Dirichlet part of the BC; when we
impose the boundary equation-of-motion we supplement it with
Neumann conditions within this boundary target space (which we
shall henceforth refer to as the D-submanifold) so that the
current conditions become \beaa k_L^{-1} j_+^L k_L & = &
\alpha\left(k_L^{-1} j_-^L k_L \right)\,
\\
k_R^{-1} j_+^R k_R & = & \alpha\left(k_R^{-1} j_-^R k_R
\right)\qquad {\rm (at}\;x=0{\rm)}. \eeaa The constant group
elements $k_L$ and $k_R$ parametrize left- and right-cosets of $H$
in $G$ and may be taken to lie in the Cartan immersion of $G/H$ in
$G$, so that the possible BCs are parametrized by $G/H\! \times
G/H$. (In fact this is true only at the level of the Lie algebras:
there is a further discrete ambiguity in the choice of $k_L,k_R$.
For details of this, and of the Cartan immersion, we refer the
reader to appendices~\ref{subsec:acbcs},~\ref{subsec:tci}.) Our
earlier results about conservation and commutation of charges and
consistency with the Poisson brackets still apply (generalized
here by twisting the currents with an inner automorphism, which
does not change the definition of the conserved charge $q_{|s|}$).

Note that when $k_L=k_R=e$ (where $e$ is the identity element in
$G$), we have $g(0)\in H$, the continuous Dirichlet boundary
parameters which determine the D-submanifold are all trivial, and
the residual symmetry is $H\!\times H$. For any $k_L,k_R$ the case
$H=G$ corresponds to the pure Neumann condition, while trivial
$H$, the pure Dirichlet condition, is inadmissible for any
non-abelian $G$.

We should point out at this stage that we have not succeeded in
finding a boundary Lagrangian for any of our mixed BCs. That is,
we have no Lagrangian of which the free variation leads to our
conditions. The Dirichlet conditions have to be imposed as
`clamped' BCs, restricting the boundary variation of $g$.

Let us now examine how much of the $G\!\times G$ symmetry
survives. We can see that the BC (\ref{bc2}) is invariant under $
k_L H k_L^{-1} \times k_R H k_R^{-1} $, and we can check that it
is precisely the charges generating this subgroup of $G\!\times G$
which are conserved on the half-line. For the global
$G$-generating charges $Q^a=\int_{-\infty}^0 j^a_0(x)\,dx$ (where
subscripts either all $L$ or all $R$ are to be understood),
consider  \beaa {d\over dt} \left(k^{-1}Qk\right) & = & k^{-1}
\left(\int_{-\infty}^0 \partial_0 j_0 \,dx\right)k
\\
& = &  k^{-1} j_1(0) k
\\ & = & {1\over 2}\left[ k^{-1} j_+ k - k^{-1} j_-
k\right]_{x=0}\\ & = & {1\over 2}\left[ \alpha(k^{-1}j_-k) -
k^{-1}j_-k \right]_{x=0}\,,\eeaa which is zero on $\h$ (only).

\subsubsection{General non-chiral BCs}

If we explore similarly the LR and RL conditions, we find that the
condition \be\label{bc3} g(0)\in g_L {G\over H}\, g_R^{-1} \ee
(where $G/H=\{\alpha(g)g^{-1}|g\in G\}$ is the Cartan immersion of
$G/H$ in $G$) leads to $g_L^{-1} j_0^L g_L = \alpha( g_R^{-1}j_0^R
g_R)$ (with $g_L,g_R$ again constant elements of $G$). If we then
apply the boundary equation-of-motion, the condition on the
currents becomes $$ g_L^{-1} j_\pm^L g_L = \alpha\left( g_R^{-1}
j_\mp^R g_R\right) \qquad{\rm at}\;x=0.$$ Unlike our chiral BCs
(which are parametrized by ${G/ H}\times {G/ H}$), these BCs are
parametrized by a single $G$ (again quotiented by a discrete
subgroup; see appendix~\ref{subsec:ancbcs}). In the specialization
$g_L=g_R$, however, the boundary is parametrized by $G/H$.

Note the inversion of the role of the dimension of $H$ in
determining the dimension of the D-submanifold, compared to the
chiral case: there (and setting $k_L=k_R=e$) we had $g(0)\in H$,
whereas here (with $g_L=g_R=e$) we have $g(0)\in G/H$. The two
extreme non-chiral cases give us nothing new: with trivial $H$ we
revert to the free, pure Neumann condition, while at the other
extreme of $H=G$ we have the pure Dirichlet condition.

As with the chiral case, we can check conservation of the
generators of the remnant of the $G\!\times G$ symmetry. This
time, because the Cartan immersion of $G/H$ is invariant under
$H_{\rm diag.}$ (the diagonal subgroup $g\mapsto hgh^{-1}$), the
surviving global symmetry is $H_{\rm diag.}$ conjugated (in
$G\!\times G$) by $(g_L,g_R)$, and we may check that \beaa {d\over
dt}\left(g_L^{-1}Q_Lg_L + g_R^{-1} Q_R g_R\right) & = & {1\over 2}
\left[ g_L^{-1}(j_+^L-j_-^L)g_L +
g_R^{-1}(j_+^R-j_-^R)g_R\right]_{x=0}
\\ & = & {1\over 2}\left[ g_R^{-1}(j_+^R-j_-^R)g_R -
\alpha(g_R^{-1}(j_+^R-j_-^R)g_R)\right]_{x=0}\eeaa which is zero
precisely on $\h\times\h$.

At this stage it is worth comparing our results with those
obtained in the Wess-Zumino-Witten model -- that is, with D-branes
on group manifolds. There, initial suggestions of a connection
with symmetric spaces \cite{kato96} were supplanted by an
understanding that the D-submanifold is actually a `twisted' or
`twined' conjugacy class \cite{felde99,stanc99}, $C_\alpha(g_0) =
\{\alpha(g)g_0 g^{-1}| g\in G\}$. This situation arises because in
the WZW model there is only one pair of currents, $j_+^L$ and
$j_-^R$, and so only one, LR, boundary condition. In our case we
have two, LR and RL, conditions, and their interplay further
requires that $\alpha(g_0)=g_0^{-1}$. But the space ${\cal M}$ of
such $g_0$ is, for the non-Grassmannian cases, precisely the
Cartan immersion $G/H$ (which is connected to the identity, so
that we can find $\sqrt{g_0}$), and
\begin{equation}\label{tcc} C_\alpha(g_0) = \{ \alpha(g) g_0 g^{-1}|g\in
G\} = \{ \alpha(g g_0^{-{1\over 2}}) (g g_0^{-{1\over 2}})^{-1} |
g\in G\} = {G\over H}\,.\end{equation} (For the Grasmannian cases,
${\cal M}$ is a union of disconnected components, the
identity-connected component being $G/H$ -- see appendix 5.
However, each of the other components is actually a translate of
the immersion of $G/H'$ for a different $H'$ \cite{AP}, so we
obtain no new BCs in this way.)

The analogous BC in our case is with $g_L=g_R=e$,  and the
residual symmetry is $H_{\rm diag.}$, necessarily preserved by any
BC utilizing a twisted conjugacy class, as in the WZW model. Note
that the $\alpha=1, H=G$ case is purely Dirichlet in our case,
whereas in the WZW model $g_0$ is unconstrained and there is still
freedom at the boundary. We expect that the non-chiral BCs should
remain integrable when a Wess-Zumino term (of arbitrary size) is
added, and we plan to explore this in future work.

Finally we note the relationship of our work to that on the
Gross-Neveu model\footnote{{\em not} the generalized chiral GN
model, which remains to be investigated} \cite{mori98b}. This
model has a single global $G=O(N)$ invariance, broken by the BC to
an $H=O(M)$ subgroup. Their boundary $S$-matrix is then diagonal.

\subsection{Remarks on quantization}

As with the bulk model, we shall assume that our results carry
through into the quantum theory: that classically conserved
charges remain conserved in the quantum theory; that charges
classically in involution do not develop ${\cal O}(\hbar^2)$
anomalies in their commutators; and that our BCs therefore lead to
quantum conservation of the charges which generate the residues of
the $G\!\times G$ symmetry. All of this leads to the expectation
that boundary scattering factorizes, so that solutions of the BYBE
provide boundary $S$-matrices.

The only technique which can give evidence for the continued
conservation of the local charges after quantization is
Goldschmidt-Witten anomaly counting \cite{gold80}. This was
carried out for the bulk case in \cite{evans97}, where for each
classical $G$ at least one non-trivial charge was found to be
necessarily conserved in the quantum theory. This was extended to
boundary models in \cite{mori98}, and used to prove quantum
conservation of the spin-3 charge for $G=SO(N)$ in \cite{macka99}
for one of our BCs. It is simple to check that for all our BCs and
for all classical $G$, each charge which necessarily survives
quantization in the bulk model also survives in the presence of
the boundary. We do not give details.

Finally, the form of our admissible D-submanifolds is not so
surprising when we remember that bulk sigma models on symmetric
spaces have particularly nice behaviour after quantization, in
that the symmetric spaces preserve their shape under
renormalization \cite{ZJ}. We would certainly expect our
D-submanifolds to behave similarly nicely.

\section{The minimal boundary $S$-matrices}
\label{sec:mbsm}

In this section we construct boundary $S$-matrices which are
minimal --- that is, which have no poles on the physical strip. We
follow the method used in the bulk case~\cite{ogiev87}, where
minimal $S$-matrices were found by solving the Yang-Baxter
equation and applying unitarity, analyticity and crossing
symmetry, and the desired pole structure then implemented using
the CDD ambiguity. In the boundary case we seek minimal boundary
$S$-matrices by making ans\"atze to solve the boundary Yang-Baxter
equation (BYBE) and applying unitarity, analyticity and the
combined crossing-unitarity relation~\cite{ghosh93}. These minimal
solutions will be used to construct PCM boundary $S$-matrices with
the appropriate pole structure in section four.

We shall make the ansatz that the boundary $S$-matrix (the
`$K$-matrix') in the defining, $N$-dimensional, vector
representation of a classical group $G$ is in one of the two forms
\be\label{form} K_1(\theta)=\rho(\theta)E \qquad{\rm and}\qquad
K_2(\theta)=\frac{\tau(\theta)}{(1- c\theta)}(I+c\theta E)\,. \ee
Here $c$ and $E$ are constants, the latter an $N\times N$ matrix;
we shall explain the $\theta$-dependent terms below. The crucial
point at this stage is the equivalent physical statement that $K$
has at most two `channels': since its matrix structure is at most
linear in rapidity $\theta$, it will decompose into one (for
$K_1$) or two (for $K_2$) projectors. This will prove sufficient
to yield a set of solutions related to the symmetric spaces in the
following

\noindent {\bf Correspondence:} for a given $G$-invariant bulk
factorized $S$-matrix ({\em i.e.\ }a solution of the YBE) in the
defining representation, the $K$-matrices of the form (\ref{form})
fall into a set of families in $1\!-\!1$ correspondence with the
set of symmetric spaces $G/H$, and each family is parametrized by
a space of admissible $E$ which is isomorphic to (possibly a
finite multiple of) the corresponding $G/H$.

Such solutions, we believe, correspond to scattering off the
boundary ground state. We would obtain solutions with many more
channels by considering scattering of bulk particles in higher
tensor representations or off higher boundary bound states, or
both. These can be obtained by fusion from our
solutions\footnote{with the exception of the $SO(N)$ spinorial
multiplets}, and the results of this paper thus lay the foundation
for future work in this direction.

The calculations of this section will necessarily, because
case-by-case and exhaustive, be rather involved. Our strategy is
to lead the reader through the implications of the BYBE,
unitarity, hermitian analyticity and crossing-unitarity, initially
culminating in a precise statement of our solutions in section
3.4. Details of the calculations are relegated to
appendices~\ref{sec:bybcuc} and~\ref{sec:apuhac}. Then in
section~\ref{subsec:constraints} we explain how our solutions are
parametrized by the symmetric spaces; details appear in
appendix~\ref{sec:apssci}.

\subsection{Calculating the boundary $S$-matrices}

Throughout the rest of this paper we shall use the following (somewhat
unconventional) notation for the bulk and boundary $S$-matrices:
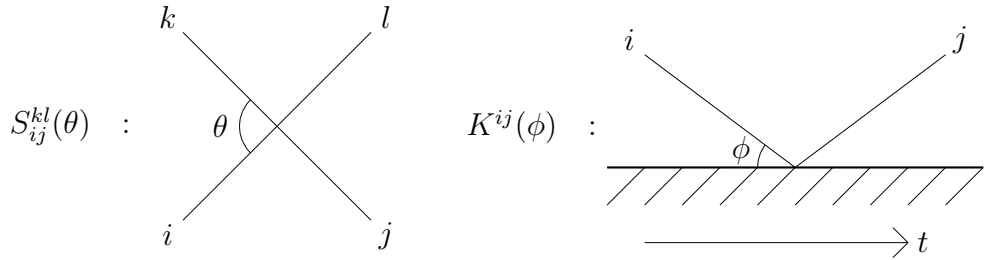
\begin{center}
\begin{pspicture}(-1,.5)(5,3.8)
\psl(1.5,3.5)(4,1)
\psl(1.5,1)(4,3.5)
\psl(2.25,2.25)(2.2601,2.35)
\psl(2.25,2.25)(2.2601,2.15)
\psl(2.2601,2.35)(2.2917,2.45)
\psl(2.2601,2.15)(2.2917,2.05)
\psl(2.2917,2.45)(2.35,2.55)
\psl(2.2917,2.05)(2.35,1.95)
\psl(2.35,2.55)(2.3964,2.6036)
\psl(2.35,1.95)(2.3964,1.8964)
\rput(2,2.25){$\theta$}
\rput(1.3,3.7){$k$}
\rput(1.3,.8){$i$}
\rput(4.2,.8){$j$}
\rput(4.2,3.7){$l$}
\rput(0,2.25){$S_{ij}^{kl}(\theta)\quad\colon$}
\end{pspicture}
\begin{pspicture}(-1,-.7)(6,2.7)
\psl(1.5,-.5)(5,-.5)
\psl(5,-.5)(4.8,-.7)
\psl(5,-.5)(4.8,-.3)
\rput(5.2,-.5){$t$}
\psline(1,.5)(6,.5)
\psl(1,0)(1.5,.5)
\psl(1.5,0)(2,.5)
\psl(2,0)(2.5,.5)
\psl(2.5,0)(3,.5)
\psl(3,0)(3.5,.5)
\psl(3.5,0)(4,.5)
\psl(4,0)(4.5,.5)
\psl(4.5,0)(5,.5)
\psl(5,0)(5.5,.5)
\psl(5.5,0)(6,.5)
\psl(1.5,2)(3.5,.5)
\psl(3.5,.5)(5.5,2)
\rput(2.78,.73){$\phi$}
\psl(3,.5)(3.0101,.6)
\psl(3.0101,.6)(3.0417,.7)
\psl(3.0417,.7)(3.1,.8)
\rput(1.3,2.2){$i$}
\rput(5.7,2.2){$j$}
\rput(0,1.05){$K^{ij}(\phi)\quad\colon$}
\end{pspicture}
\end{center}
where $\theta$ is the rapidity difference between the two in-coming
particles which scatter in the first diagram and $\phi$ is the rapidity
of the in-coming particle reflecting in the second.

We consider the BYBE for two particles in the vector
representation \bea \label{eqn:bybe} S_{ij}^{kl}(\theta-\phi)( I_{jm}
\otimes K^{ln}(\theta)) S_{mo}^{np}(\theta+\phi)( I_{oq} \otimes
K^{pr}(\phi))= \phantom{MMMMMMMM} \nonumber \\ \phantom{MMMMMMMM}
(I_{ij} \otimes K^{kl}(\phi))S_{jm}^{ln}(\theta+\phi) (I_{mo}
\otimes K^{np}(\theta)) S_{oq}^{pr}(\theta-\phi) \eea
($\theta$ and $\phi$ are now the rapidities of the two particles.)
We attempt to find solutions of the form (equivalent to
(\ref{form}))\be K_1(\theta)=\rho(\theta)E \qquad{\rm and}\qquad
K_2(\theta)= \tau(\theta) \left(
P^--\left[\frac{h}{ci\pi}\right]P^+\right) \ee where
$\rho(\theta)$ and $\tau(\theta)$ are scalar prefactors, $h$ is
the dual Coxeter number of the group, $c$ is a parameter and
$$P^{\pm}=\frac{1}{2}\left(I\pm E\right) \qquad\qquad
[x]=\frac{\theta+\frac{i\pi x}{h}}{\theta-\frac{i\pi x}{h}}\,,$$
where $I$ is the identity matrix and $E$ is a general square
matrix of the same size\footnote{The apparently circuitous
involvement of $i\pi/h$ ensures that $x$ is an integer, as is
usual in the literature.}. We find that constraints are imposed on
the scalar prefactors and the matrix $E$ by the BYBE and
unitarity, analyticity and crossing-unitarity. These constraints
are in general dependent on the choice of classical group,
$SU(N)$, $SO(N)$ or $Sp(N)$, but in all cases involving
$K_2(\theta)$ we find that $E^2=1$, so that $P^{\pm}$ are
projectors, as the notation suggests.

We note that the constraints imposed on the scalar prefactors will
allow us to find them only up to the usual CDD ambiguity. This
freedom is then further restricted by taking the $K$-matrices to
be minimal -- that is, taking $\rho(\theta)$ and $\tau(\theta)$ to
have no poles on the physical strip. We shall also require
$\tau(\theta)$ to have a zero at $\theta=\frac{1}{c}$ so that
$K_2(\theta)$ is finite at this point, since we shall find that
$\frac{1}{c}$ can lie in the physical strip.

For the case of $SU(N)$ the vector representation is not
self-conjugate. This allows us to consider the situation where a
particle scattering off the boundary returns as an anti-particle
(for an analogous situation see \cite{gand99}). In this case the
$K$-matrix, $K^{i\bar{j}}(\theta)$, must satisfy a version of the
BYBE which for convenience we shall refer to as the `conjugated'
BYBE, \bea \label{Conj}S_{ij}^{kl}(\theta-\phi)( I_{jm} \otimes
K^{l\bar{n}}(\theta)) S_{m\bar{o}}^{\bar{n}p}(\theta+\phi)(
I_{\bar{o}\bar{q}} \otimes K^{p\bar{r}}(\phi))= \phantom{MMMMMMMM}
\nonumber \\ \phantom{MMMMMMMM} (I_{ij} \otimes
K^{k\bar{l}}(\phi))S_{j\bar{m}}^{\bar{l}n}(\theta+\phi)
(I_{\bar{m}\bar{o}} \otimes K^{n\bar{p}}(\theta))
S_{\bar{o}\bar{q}}^{\bar{p}\bar{r}}(\theta-\phi)\,. \eea We shall
find that only $K_1(\theta)$ gives a solution of the conjugated
BYBE. Analyticity and unitarity become more subtle in this case.
Henceforth we shall refer to this case as the `conjugating' case for
the $SU(N)$ model.

Even in the `non-conjugating' $SU(N)$ model, where the representation
is preserved under scattering by the boundary, the fact that the vector
representation is not self-conjugate leads to subtleties in the
calculations. In addition to the standard BYBE of~(\ref{eqn:bybe}) there
is also the possibility that one or both of the in-coming particles in
the BYBE scattering process is the conjugate vector particle. Thus we
need to introduce a minimal $K$-matrix, $\bar{K}(\theta)$, describing
this scattering. We take $\bar{K}(\theta)$ to have the same one or two
channel form as $K(\theta)$, that is
\be \bar{K}_1(\theta)=\bar{\rho}(\theta)F \qquad{\rm or}\qquad
\bar{K}_2(\theta)=\frac{\bar{\tau}(\theta)}{(1- d\theta)}(I+d\theta
F)\,, \ee
where we have a new set of scalar prefactors and constants,
$\bar{\rho}(\theta)$, $\bar{\tau}(\theta)$, $d$ and $F$.

At this point we introduce the diagrammatic algebra used in the
calculations we perform. We represent the matrices $I$, $J$, $E$ and $F$,
where $J$ is the symplectic form matrix, in the following
way: $$I\colon\; \one \qquad J\colon\;\onjr \qquad E\colon\;\ons \qquad
F\colon\;\onsc \qquad \left(J^t\colon\;\onjl\right)$$ (Thus from the
properties of $J$ we have $\onjr=-(\onjl)$ and $\onjr\onjl=\one=\onjl\onjr$.)
Matrix multiplication is achieved by concatenation of the
diagrams, for example $$JIJ^tE\colon\; \onjr \one \onjl \ons =
\onjr \onjl \ons = \ons$$ Using this diagrammatic algebra we can
rewrite the $K$-matrices of (\ref{form}) as
\be
K_1(\theta)=\rho(\theta)\ons\qquad\qquad
K_2(\theta)=\frac{\tau(\theta)}{(1-
c\theta)}\left(\one+c\theta\ons\right) \ee This diagrammatic
description is used in the calculations presented in the
appendices, and makes them much clearer.

We substitute either of these $K$-matrices into the BYBE, taking
the minimal $S$-matrix to be that of the bulk PCM for a particular
$G$, derived in \cite{ogiev87} up to a few minor inconsistencies
which we have corrected. This yields constraints (dependent on $G$
and on whether we take $K_1$ or $K_2$) that $E$ (and $F$) must satisfy
in order for $K_i$ to be a solution of the BYBE. In fact the
constraints from crossing-unitarity on the $K_i$ are identical to those
imposed by the BYBE, therefore we shall delay presentation of the
constraints until later. (The reader is referred to
appendix~\ref{subsec:bybecal} for details of the BYBE calculations.)
Instead we go on to consider unitarity, analyticity and
crossing-unitarity.

\subsection{Unitarity and Analyticity}
\label{subsec:uha}

\subsubsection{The non-conjugating cases}
\label{ssubsec:uhanc}

For the non-conjugating cases the $K$-matrices are required to
satisfy the conditions of unitarity~\cite{ghosh93} and hermitian
analyticity~\cite{mira99}
\be
K(\theta)K(-\theta)=I\qquad\qquad\qquad
K(\theta)=\left(K(-\theta^{\ast})\right)^{\dag}\,. \ee
Substituting $K_1$ we obtain
$$\rho(\theta)\rho(-\theta)\ons\ons=\one\qquad\qquad\rho(\theta)\ons=
\rho(-\theta^{\ast})^{\ast}(\ons)^{\dag}\,,$$ These matrix
equations are equivalent to
$$\rho(\theta)\rho(-\theta)=\frac{1}{\alpha}\qquad\qquad\rho(\theta)
=\beta\rho(-\theta^{\ast})^{\ast}$$
$$\ons\ons=\alpha\one\quad\qquad\qquad(\ons)^{\dag}=\beta\ons$$
where $\alpha$ and $\beta$ are constants with $\beta\in U(1)$.
Recalling $K_1(\theta)$, we see that we have the freedom in our
definition of $\rho(\theta)$ and $\ons$ to set $\beta=1$ and
ensure $\alpha\in U(1)$, so that the constraints imposed on the
matrix $\ons$ by unitarity and analyticity become
\be\label{mbsmun} (\ons)^{\dag}=\ons\quad{\rm
and}\quad\ons\ons=\alpha\one\quad{\rm where}\ \;\alpha\in U(1).
\ee

Similarly for $K_2$ we obtain
$$\frac{\tau(\theta)\tau(-\theta)}{(1-c^2\theta^2)}\left(\one
-c^2\theta^2\ons\ons\right)=\one$$
$$\frac{\tau(\theta)}{(1-c\theta)} \left( \one +c\theta \ons
\right )=\frac{\tau(-\theta^{\ast})^{\ast}}{(1+c^{\ast}\theta)}
\left( \one -c^{\ast}\theta(\ons)^{\dag}\right)$$ These are
equivalent to \beaa
\tau(\theta)\tau(-\theta)=\frac{(1-c^2\theta^2)}{(1-\gamma
c^2\theta^2)}\qquad\qquad\ons\ons=\gamma \one \\
\frac{\tau(\theta)}{(1-c\theta)}
=\frac{\tau(-\theta^{\ast})^{\ast}}{(1+c^{\ast}\theta)}
\qquad\qquad c\ons=-c^{\ast}(\ons)^{\dag} \eeaa If we consider the
expression for $K_2(\theta)$, we see that we have the freedom to
set $c$ to be purely imaginary and choose $\gamma\in U(1)$. Then
the constraints on the matrix $\ons$ become
$$(\ons)^{\dag}=\ons\quad{\rm
and}\quad\ons\ons=\gamma\one\quad{\rm where}\ \;\gamma\in U(1)$$
as was the case for $K_1$. In fact we find that the parameters
$\alpha$ and $\gamma$ cannot be freely chosen in $U(1)$; the only
choice consistent with the hermiticity of $\ons$ is that they are
equal to $1$ (see appendix~\ref{subsec:hanc}). Consequently we
find that for both $K_1$ and $K_2$, unitarity and analyticity
impose
\be
(\ons)^{\dag}=\ons\quad{\rm and}\quad\ons\ons=\one\,. \ee The
corresponding constraints imposed on $\rho(\theta)$ and
$\tau(\theta)$ are \bea \rho(\theta)=\rho(-\theta^{\ast})^{\ast}&
\qquad & \rho(\theta)\rho(-\theta)=1 \\
\tau(\theta)=\tau(-\theta^{\ast})^{\ast}& \qquad &
\tau(\theta)\tau(-\theta)=1\,. \eea
(We note that similar conditions are imposed on $\bar{\rho}(\theta)$,
$\bar{\tau}(\theta)$ and $\onsc$.)

\subsubsection{$SU(N)$-conjugating}
\label{ssubsec:uhac}

In the case of the conjugated BYBE (\ref{Conj}), which we call
`$SU(N)$-conjugating', unitarity and analyticity can no longer be
applied as straightforwardly. The reason is that we are no longer
dealing with a $K$-matrix that is an endomorphism of the vector
representation space $V$, but rather with $K(\theta):V \to
\bar{V}$. In order to apply our conditions we must introduce
$K'(\theta):\bar{V} \to V$ and consider the space $V\oplus\bar{V}$
on which the endomorphism
$$\tilde{K}(\theta)=\left(\begin{array}{cc} 0 & K'(\theta)
\\ K(\theta) & 0
\end{array}\right)\qquad {\rm acts.}$$
We can then apply analyticity and unitarity to $\tilde{K}(\theta)$
which yields
\be
K(\theta)=K'(-\theta^{\ast})^{\dag}\qquad\qquad\qquad
K(\theta)K'(-\theta)=I\,. \ee

The conjugated BYBE that $K(\theta)$ and $K'(\theta)$ must satisfy
allows only the $K_1(\theta)$ form for both. Thus we have
$$K(\theta)=\rho(\theta)\ons\qquad{\rm and}\qquad
K'(\theta)=\omega(\theta)\onsc$$ where $\rho(\theta)$ and
$\omega(\theta)$ are scalar prefactors, whilst $\ons= E$ and
$\onsc\equiv F$ are constant matrices. Substituting these into our
conditions yields \bea\label{mbsmuc1}
\rho(\theta)=\alpha\omega(-\theta^{\ast})^{\ast}\qquad\qquad\qquad
\rho(\theta)\omega(-\theta)=\frac{1}{\beta} \nonumber \\
\alpha\ons=(\onsc)^{\dag}\qquad\qquad\quad\qquad\ons\onsc=\beta\one
\phantom{n}\,. \eea

There is enough rescaling freedom in splitting $K(\theta)$ and
$K'(\theta)$ into scalar prefactors and matrix parts that we can
consistently set $\alpha\!=\!\beta\!=\!1$ and $\det{E}\!=\!1
\Longleftrightarrow\det{F}\!=\!1$ (details in
appendix~\ref{subsec:hac}). This leaves us with
\bea E=F^{\dag},\quad & EF=I, & \det{F}=\det{E}=1, \nonumber \\
\rho(\theta)=\omega(-\theta^{\ast})^{\ast} & {\rm and} &
\quad\rho(\theta)\omega(-\theta)=1. \eea

\subsection{Crossing-Unitarity}
\subsubsection{The non-conjugating cases}

The boundary $S$-matrices must also satisfy
crossing-unitarity~\cite{ghosh93}, which in our notation is
\be
K^{ij}(\frac{i\pi}{2}-\theta)=S_{\bar{j}k}^{i\bar{l}}(2\theta)K^{\bar{l}
\bar{k}}(\frac{i\pi}{2}+\theta)\,. \ee Substituting the $K_i$
into the crossing-unitarity equation gives constraints on $\ons$, $c$,
$d$ and the scalar prefactors.  We tabulate these below (details in
appendix ~\ref{subsec:cucal}). Note that in our diagrammatic algebra
$Tr(E)$ is represented by the symbol $\trs$.
\begin{center}
\begin{tabular}{|c|c|}\hline
$\begin{array}{c}
{\rm Group} \\
K_1(\theta)/K_2(\theta)
\end{array}$ & Constraints \\ \hline
$SU(N)$ & $\trs=0$, \\
$K_1(\theta)$ & $\vs_{\vs}\rho(i\pi\!-\!\theta)\rho(i\pi\!+\!\theta)=
{\displaystyle\frac{(\theta\!-\!\frac{i\pi}{h})(\theta\!+\!\frac{i\pi}
{h})}{\theta^2}}\vs_{\vs}$ \\ \hline
$SU(N)$ & $\trs=\frac{ih}{\pi}\left(\frac{1}{c}\!+\!\frac{1}{d}\right)$,
\\
$K_2(\theta)$ & $\vse_{\vse}\tau(i\pi\!-\!\theta)\tau(i\pi\!+\!\theta)=
{\displaystyle\frac{(\theta\!-\!\frac{i\pi}{h})(\theta\!+\!\frac{i\pi}
{h})(\theta\!-\!i\pi\!+\!\frac{1}{c})(\theta\!+\!i\pi\!-\!\frac{1}{c})}
{(\theta\!-\!i\pi)(\theta\!+\!i\pi)(\theta\!-\!\frac{1}{d})(\theta\!+\!
\frac{1}{d})}}\vse_{\vse}$ \\ \hline
$SO(N)$ & $\trs=0$\quad{\small and either} \\
$K_1(\theta)$ & $\begin{array}{c}
(\ons)^t=\ons,\quad\frac{\rho(i\pi/2-\theta)}{\rho(i\pi/2+\theta)}=
\sigma_o(2\theta) \\
{\rm\small or} \\
(\ons)^t=-(\ons),\quad\frac{\rho(i\pi/2-\theta)}{\rho(i\pi/2+\theta)}=
-[1]\sigma_o(2\theta)
\end{array}$ \\ \hline
$SO(N)$ & $\trs=\frac{2ih}{c\pi},\quad (\ons)^t=\ons,\quad\frac{\tau(i
\pi/2-\theta)}{\tau(i\pi/2+\theta)}=[\frac{h}{2}]\left[\frac{h}{ci\pi}\!
-\!\frac{h}{2}\right]\sigma_o(2\theta)$ \\
$K_2(\theta)$ & $\begin{array}{c}
{\rm\small or} \\
(\ons)^t=-(\ons),\quad\frac{\tau(i\pi/2-\theta)}{\tau(i\pi/2+\theta)}=
-[1][\frac{h}{2}]\left[\frac{h}{ci\pi}\!-\!\frac{h}{2}\right]\sigma_o(2
\theta)
\end{array}$ \\ \hline
$S\!p(N)$ & $\trs=0$\quad{\small and either} \\
$K_1(\theta)$ & $\begin{array}{c}
\onjr(\ons)^t\onjr=\ons,\quad\frac{\rho(i\pi/2-\theta)}{\rho(i\pi/2+
\theta)}=\sigma_p(2\theta) \\
{\rm\small or} \\
\onjr(\ons)^t\onjr=-(\ons),\quad\frac{\rho(i\pi/2-\theta)}{\rho(i\pi/2+
\theta)}=-[1]\sigma_p(2\theta)
\end{array}$ \\ \hline
$Sp(N)$ & $\onjr(\ons)^t\onjr=\ons,\quad\frac{\tau(i\pi/2-\theta)}{
\tau(i\pi/2+\theta)}=[\frac{h}{2}]\left[\frac{h}{ci\pi}\!-\!\frac{h}{2}
\right]\sigma_p(2\theta)$ \\
$K_2(\theta)$ & $\begin{array}{c}
{\rm\small or} \\
\trs=\frac{2ih}{c\pi},\quad \onjr(\ons)^t\onjr=-\ons, \\
\frac{\tau(i\pi/2-\theta)}{\tau(i\pi/2+\theta)}=-[1][\frac{h}{2}]\left[
\frac{h}{ci\pi}\!-\!\frac{h}{2}\right]\sigma_p(2\theta)
\end{array}$ \\ \hline
\end{tabular}
\end{center}
\vspace{1ex} The functions $\sigma_o$ and $\sigma_p$ are scalar
prefactors for the bulk $S$-matrices (see appendix~\ref{sec:bybcuc}).

\subsubsection{$SU(N)$-conjugating}

The $SU(N)$-conjugating $K$-matrix must satisfy the crossing-unitarity
equation
\be
K^{i\bar{j}}(\frac{i\pi}{2}-\theta)=S_{jk}^{il}(2\theta)K^{l\bar{k}}
(\frac{i\pi}{2}+\theta)\,. \ee
From this we obtain the constraints (see appendix~\ref{sec:bybcuc})
\be\label{eqn:rho}\frac{\rho(\frac{i\pi}{2}-\theta)}{\rho(\frac{i\pi}
{2}+\theta)}=\left\{\begin{array}{ll} \sigma_u(2\theta) & (\ons)^t=\ons \\
-[1]\sigma_u(2\theta) & (\ons)^t=-(\ons).
\end{array}\right.\ee
Similarly, the crossing-unitarity equation for $K'(\theta)$ yields
identical constraints on $\omega(\theta)$ and $\onsc$. Since
$\ons\onsc=\ons$ the signs in $(\ons)^t=\pm(\ons)$ and $(\onsc)^t=\pm(
\onsc)$ must coincide. From this we have
\be\frac{\rho(\frac{i\pi}{2}-\theta)}{\rho(\frac{i\pi}{2}+\theta)}=
\frac{\omega(\frac{i\pi}{2}-\theta)}{\omega(\frac{i\pi}{2}+\theta)},\ee
which, together with the other constraints on these scalar prefactors,
is enough to show we should take $\rho(\theta)\!=\!\omega(\theta)$.

Now $F$ and $\omega(\theta)$ are completely fixed by $E$ and
$\rho(\theta)$, and the boundary $S$-matrix for $V\oplus\bar{V}$ is
\be
\tilde{K}(\theta)=\rho(\theta)\left(\begin{array}{cc}
0 & E^{\dag}\\
E & 0
\end{array}\right)
\ee subject to the following constraints, along with~(\ref{eqn:rho}):
\bea E^{\dag}E=I\qquad&{\rm and}&\qquad\det{E}=1
\nonumber \\ \rho(\theta)=\rho(-\theta^{\ast})^{\ast}\quad&{\rm
and}&\qquad\rho(\theta)\rho(-\theta)=1\,. \eea

\subsection{The boundary $S$-matrices}
\label{subsec:bsm}

We have obtained a series of constraints on $\ons$ and on
$\rho(\theta)$ and $\tau(\theta)$ which must be satisfied if the
proposed $K_1(\theta)$ and $K_2(\theta)$ are to be boundary
$S$-matrices. The constraints on the scalar prefactors enable us
to determine them exactly for each $G$, providing we assume some
extra minimality conditions, namely that they should be
meromorphic functions of $\theta$ with no poles on the physical
strip Im$\,\theta\in[0,\frac{\pi}{2}]$. Having calculated the
scalar prefactors (we do not include details of the calculations,
as the reader can simply check the results if required) we obtain
the boundary $S$-matrices below.

We note that in the case of $K_2(\theta)$ there is also the
possibility that the pole at $\theta=\frac{1}{c}$ lies on the
physical strip and so the scalar prefactor $\tau(\theta)$ may be
required to have a zero at this point. In fact we shall find that
one of $\pm\frac{1}{c}$ always lies on the physical strip, and the
expressions given below for $K_2(\theta)$, for each case, are
valid when $\frac{1}{c}$ lies on the physical strip. When
$-\frac{1}{c}$ lies on the physical strip instead, the correct
expressions for the minimal $K_2$-matrices can be obtained by the
interchange $c\leftrightarrow -c,\,\ons\leftrightarrow -\ons$ (together
with $d\leftrightarrow -d$ for $SU(N)$). (Note that this leaves
$c\,\ons$ unchanged.) In section~\ref{sec:pcmbsm} we shall add CDD
factors making the PCM boundary $S$-matrices invariant under this
interchange.

\subsubsection{$SU(N)$}

The minimal boundary $S$-matrices for $SU(N)$ are
\be
K_1(\theta)=\frac{\Gamma\left(\frac{\theta}{2i\pi}+
\frac{1}{2}+\frac{1}{2h}\right)\Gamma\left(\frac{-\theta}{2i\pi}+\frac{
1}{2}\right)}{\Gamma\left(\frac{-\theta}{2i\pi}+\frac{1}{2}+\frac{1}{2h}
\right)\Gamma\left(\frac{\theta}{2i\pi}+\frac{1}{2}\right)}\,\ons\qquad
{\rm with}\ \trs=0 \ee where $\Gamma$ is the gamma function, and
\be
K_2(\theta)=\frac{-1}{(1-c\theta)}\frac{\Gamma\left(\frac{\theta}
{2i\pi}+\frac{1}{2}+\frac{1}{2h}\right)\Gamma\left(\frac{-\theta}{2i\pi}
\right)\Gamma\left(\frac{\theta}{2i\pi}+\frac{1}{2i\pi c}\right)\Gamma
\left(\frac{-\theta}{2i\pi}+\frac{1}{2}+\frac{1}{2i\pi d}\right)}{\Gamma
\left(\frac{-\theta}{2i\pi}+\frac{1}{2}+\frac{1}{2h}\right)\Gamma\left(
\frac{\theta}{2i\pi}\right)\Gamma\left(\frac{-\theta}{2i\pi}+\frac{1}{
2i\pi c}\right)\Gamma\left(\frac{\theta}{2i\pi}+\frac{1}{2}+\frac{1}{
2i\pi d}\right)}(\one+c\theta\ons)
\ee
with $\trs=\frac{ih}{\pi}(\frac{1}{c}\!+\!\frac{1}{d})$.

\noindent (Note that in the limit $c,d\to\infty$, with
$\frac{cd}{c+d}\trs$ fixed, $K_2(\theta)\to K_1(\theta)$, as we would
expect.)

\subsubsection{$SO(N)$}

There are two types of minimal boundary $S$-matrix, corresponding
to $E$ symmetric or antisymmetric, for each of the $K_1(\theta)$
and $K_2(\theta)$ forms in the case $G\!=\!SO(N)$. (The symmetric
$K_2(\theta)$ type was investigated in \cite{macka95} and, for
$M=1$, in \cite{ghosh94}. The $K_2(\theta)$ type with
antisymmetric $E$ has been considered in~\cite{moric01}.) The
minimal boundary $S$-matrices of the $K_1(\theta)$ form are
\be
K_1(\theta)=\frac{\Gamma\left(\frac{\theta}{2i\pi}+\frac{3}{4}\right)
\Gamma\left(\frac{-\theta}{2i\pi}+\frac{1}{2}\right)\Gamma\left(\frac{
\theta}{2i\pi}+\frac{1}{2}+\frac{1}{2h}\right)\Gamma\left(\frac{-\theta}
{2i\pi}+\frac{1}{4}+\frac{1}{2h}\right)}{\Gamma\left(\frac{-\theta}{
2i\pi}+\frac{3}{4}\right)\Gamma\left(\frac{\theta}{2i\pi}+\frac{1}{2}
\right)\Gamma\left(\frac{-\theta}{2i\pi}+\frac{1}{2}+\frac{1}{2h}\right)
\Gamma\left(\frac{\theta}{2i\pi}+\frac{1}{4}+\frac{1}{2h}\right)}\,\ons
\ee with $(\ons)^t=\ons$, $\trs=0$, and
\be
K_1(\theta)=\frac{\Gamma\left(\frac{\theta}{2i\pi}+\frac{3}{4}\right)
\Gamma\left(\frac{-\theta}{2i\pi}+\frac{1}{2}\right)\Gamma\left(\frac{
\theta}{2i\pi}+\frac{1}{2}+\frac{1}{2h}\right)\Gamma\left(\frac{-\theta}
{2i\pi}+\frac{3}{4}+\frac{1}{2h}\right)}{\Gamma\left(\frac{-\theta}{
2i\pi}+\frac{3}{4}\right)\Gamma\left(\frac{\theta}{2i\pi}+\frac{1}{2}
\right)\Gamma\left(\frac{-\theta}{2i\pi}+\frac{1}{2}+\frac{1}{2h}\right)
\Gamma\left(\frac{\theta}{2i\pi}+\frac{3}{4}+\frac{1}{2h}\right)}\,\ons
\ee with $(\ons)^t=-(\ons)$, $\left(\Rightarrow\trs=0\right)$. Those of
the $K_2(\theta)$ form are
\bea
K_2(\theta)=\frac{-1}{(1-c\theta)}\frac{\Gamma\left(\frac{\theta}{2i\pi}
+\frac{3}{4}\right)\Gamma\left(\frac{-\theta}{2i\pi}\right)\Gamma\left(
\frac{\theta}{2i\pi}+\frac{1}{2}+\frac{1}{2h}\right)\Gamma\left(\frac{
-\theta}{2i\pi}+\frac{1}{4}+\frac{1}{2h}\right)}{\Gamma\left(\frac{
-\theta}{2i\pi}+\frac{3}{4}\right)\Gamma\left(\frac{\theta}{2i\pi}
\right)\Gamma\left(\frac{-\theta}{2i\pi}+\frac{1}{2}+\frac{1}{2h}\right)
\Gamma\left(\frac{\theta}{2i\pi}+\frac{1}{4}+\frac{1}{2h}\right)}
\nonumber \\ \times\frac{\Gamma\left(\frac{\theta}{2i\pi}+\frac{1}{2i\pi
c}\right)
\Gamma\left(\frac{-\theta}{2i\pi}+\frac{1}{2}+\frac{1}{2i\pi
c}\right)} {\Gamma\left(\frac{-\theta}{2i\pi}+\frac{1}{2i\pi
c}\right)\Gamma
\left(\frac{\theta}{2i\pi}+\frac{1}{2}+\frac{1}{2i\pi
c}\right)}(\one+c \theta\ons) \eea with $(\ons)^t=\ons$,
$c\trs=\frac{2ih}{\pi}$, and
\bea
K_2(\theta)=\frac{-1}{(1-c\theta)}\frac{\Gamma\left(\frac{\theta}{2i\pi}
+\frac{3}{4}\right)\Gamma\left(\frac{-\theta}{2i\pi}\right)\Gamma\left(
\frac{\theta}{2i\pi}+\frac{1}{2}+\frac{1}{2h}\right)\Gamma\left(\frac{
-\theta}{2i\pi}+\frac{3}{4}+\frac{1}{2h}\right)}{\Gamma\left(\frac{
-\theta}{2i\pi}+\frac{3}{4}\right)\Gamma\left(\frac{\theta}{2i\pi}
\right)\Gamma\left(\frac{-\theta}{2i\pi}+\frac{1}{2}+\frac{1}{2h}\right)
\Gamma\left(\frac{\theta}{2i\pi}+\frac{3}{4}+\frac{1}{2h}\right)}
\nonumber \\ \times\frac{\Gamma\left(\frac{\theta}{2i\pi}+\frac{1}{2i\pi
c}\right)\Gamma\left(\frac{-\theta}{2i\pi}+\frac{1}{2}+\frac{1}{2i\pi c}
\right)} {\Gamma\left(\frac{-\theta}{2i\pi}+\frac{1}{2i\pi c}\right)
\Gamma\left(\frac{\theta}{2i\pi}+\frac{1}{2}+\frac{1}{2i\pi c}\right)}
(\one+c \theta\ons) \eea with $(\ons)^t=-(\ons)$, $\left(\Rightarrow\trs
=0\right)$.

\noindent (Note that in the limit $c\to\infty$, with $c\trs$
fixed, the symmetric (respectively antisymmetric) $K_2(\theta)$
tends to the symmetric (respectively antisymmetric) $K_1(\theta)$,
again as expected.)

\subsubsection{$S\!p(N)$}

For $S\!p(N)$ there are again two minimal solutions of the form
$K_1(\theta)$:
\be
K_1(\theta)=\frac{\Gamma\left(\frac{\theta}{2i\pi}+\frac{3}{4}\right)
\Gamma\left(\frac{-\theta}{2i\pi}+\frac{1}{2}\right)\Gamma\left(\frac{
\theta}{2i\pi}+\frac{1}{2}+\frac{1}{2h}\right)\Gamma\left(\frac{-\theta}
{2i\pi}+\frac{1}{4}+\frac{1}{2h}\right)}{\Gamma\left(\frac{-\theta}{
2i\pi}+\frac{3}{4}\right)\Gamma\left(\frac{\theta}{2i\pi}+\frac{1}{2}
\right)\Gamma\left(\frac{-\theta}{2i\pi}+\frac{1}{2}+\frac{1}{2h}\right)
\Gamma\left(\frac{\theta}{2i\pi}+\frac{1}{4}+\frac{1}{2h}\right)}\,\ons
\ee with $\onjr(\ons)^t\onjr=\ons$, $\left(\Rightarrow\trs=0\right)$,
and
\be
K_1(\theta)=\frac{\Gamma\left(\frac{\theta}{2i\pi}+\frac{3}{4}\right)
\Gamma\left(\frac{-\theta}{2i\pi}+\frac{1}{2}\right)\Gamma\left(\frac{
\theta}{2i\pi}+\frac{1}{2}+\frac{1}{2h}\right)\Gamma\left(\frac{-\theta}
{2i\pi}+\frac{3}{4}+\frac{1}{2h}\right)}{\Gamma\left(\frac{-\theta}{
2i\pi}+\frac{3}{4}\right)\Gamma\left(\frac{\theta}{2i\pi}+\frac{1}{2}
\right)\Gamma\left(\frac{-\theta}{2i\pi}+\frac{1}{2}+\frac{1}{2h}\right)
\Gamma\left(\frac{\theta}{2i\pi}+\frac{3}{4}+\frac{1}{2h}\right)}\,\ons
\ee with $\onjr(\ons)^t\onjr=-(\ons)$, $\trs=0$. The two minimal
boundary $S$-matrices of the form $K_2(\theta)$ are
\bea
K_2(\theta)=\frac{-1}{(1-c\theta)}\frac{\Gamma\left(\frac{\theta}{2i\pi}
+\frac{3}{4}\right)\Gamma\left(\frac{-\theta}{2i\pi}\right)\Gamma\left(
\frac{\theta}{2i\pi}+\frac{1}{2}+\frac{1}{2h}\right)\Gamma\left(\frac{-
\theta}{2i\pi}+\frac{1}{4}+\frac{1}{2h}\right)}{\Gamma\left(\frac{-
\theta}{2i\pi}+\frac{3}{4}\right)\Gamma\left(\frac{\theta}{2i\pi}\right)
\Gamma\left(\frac{-\theta}{2i\pi}+\frac{1}{2}+\frac{1}{2h}\right)\Gamma
\left(\frac{\theta}{2i\pi}+\frac{1}{4}+\frac{1}{2h}\right)} \nonumber \\
\times\frac{\Gamma\left(\frac{\theta}{2i\pi}+\frac{1}{2i\pi c}\right)
\Gamma\left(\frac{-\theta}{2i\pi}+\frac{1}{2}+\frac{1}{2i\pi c}\right)}{
\Gamma\left(\frac{-\theta}{2i\pi}+\frac{1}{2i\pi c}\right)\Gamma\left(
\frac{\theta}{2i\pi}+\frac{1}{2}+\frac{1}{2i\pi c}\right)}(\one+c\theta
\ons) \eea with $\onjr(\ons)^t\onjr=\ons$, $\left(\Rightarrow\trs=0
\right)$, and
\bea
K_2(\theta)=\frac{-1}{(1-c\theta)}\frac{\Gamma\left(\frac{\theta}{2i\pi}
+\frac{3}{4}\right)\Gamma\left(\frac{-\theta}{2i\pi}\right)\Gamma\left(
\frac{\theta}{2i\pi}+\frac{1}{2}+\frac{1}{2h}\right)\Gamma\left(\frac{-
\theta}{2i\pi}+\frac{3}{4}+\frac{1}{2h}\right)}{\Gamma\left(\frac{-
\theta}{2i\pi}+\frac{3}{4}\right)\Gamma\left(\frac{\theta}{2i\pi}\right)
\Gamma\left(\frac{-\theta}{2i\pi}+\frac{1}{2}+\frac{1}{2h}\right)\Gamma
\left(\frac{\theta}{2i\pi}+\frac{3}{4}+\frac{1}{2h}\right)} \nonumber \\
\times\frac{\Gamma\left(\frac{\theta}{2i\pi}+\frac{1}{2i\pi c}\right)
\Gamma\left(\frac{-\theta}{2i\pi}+\frac{1}{2}+\frac{1}{2i\pi c}\right)}{
\Gamma\left(\frac{-\theta}{2i\pi}+\frac{1}{2i\pi c}\right)\Gamma\left(
\frac{\theta}{2i\pi}+\frac{1}{2}+\frac{1}{2i\pi c}\right)}(\one+c\theta
\ons) \eea with $\onjr(\ons)^t\onjr=-(\ons)$, $c\trs=\frac{2ih}{\pi}$.

\noindent (As $c\to\infty$, with $c\trs$ fixed, the $K_2(\theta)$
with $\onjr(\ons)^t\onjr=\pm(\ons)$ tends to the $K_1(\theta)$
with the respective property, as expected.)

\subsubsection{$SU(N)$-conjugating}

Here it is only $K_1(\theta)$ that provides valid boundary
$S$-matrices. There are two minimal possibilities, with symmetric
and antisymmetric $\ons$,
\be
K_1(\theta)=\frac{\Gamma\left(\frac{\theta}{2i\pi}+\frac{1}{4}\right)
\Gamma\left(\frac{-\theta}{2i\pi}+\frac{1}{4}+\frac{1}{2h}\right)}{
\Gamma\left(\frac{-\theta}{2i\pi}+\frac{1}{4}\right)\Gamma\left(\frac{
\theta}{2i\pi}+\frac{1}{4}+\frac{1}{2h}\right)}\,\ons\qquad{\rm
with}\ (\ons)^t=\ons, \;\;{\rm and} \ee
\be
K_1(\theta)=\frac{\Gamma\left(\frac{\theta}{2i\pi}+\frac{1}{4}\right)
\Gamma\left(\frac{-\theta}{2i\pi}+\frac{3}{4}+\frac{1}{2h}\right)}{
\Gamma\left(\frac{-\theta}{2i\pi}+\frac{1}{4}\right)\Gamma\left(\frac{
\theta}{2i\pi}+\frac{3}{4}+\frac{1}{2h}\right)}\,\ons\qquad{\rm
with}\ (\ons)^t=-(\ons)\,. \ee We have been unable to make contact
between our solutions and a rational limit of the trigonometric
solutions in \cite{gand99}.

\subsection{Constraints on $E$: the symmetric-space correspondence}
\label{subsec:constraints}

We now turn our attention to the constraints imposed on the
matrices $E$. We recall that \be\label{conss} E^{\dag}=E\quad{\rm
and}\quad E^2=I \ee were to be imposed in all
 cases, except that of $SU(N)$-conjugating, due to unitarity and
 analyticity. For the case of
$SU(N)$-conjugating (\ref{conss}) is replaced by
\be
E^{\dag}E=I\quad{\rm and}\quad\det{E}=1\,. \ee The further
constraints particular to the different groups were
\begin{center}
\begin{tabular}{|c|c|c|}\hline
Group & $K_1(\theta)$ & $K_2(\theta)$ \\ \hline
$SU(N)$ & Tr$(E)=0$ & $\vsu_{\vsu}$ Tr$(E)=\frac{ih}{\pi}(\frac{1}{c}\!+
\!\frac{1}{d})$ $\vsu_{\vsu}$ \\ \hline
$SO(N)$ & Tr$(E)=0$, & Tr$(E)=\frac{2ih}{c\pi}$,\quad or\quad Tr$(E)=0$,
\\
& $E^t=\pm E$ & \,$E^t=E$\qquad\qquad\,$E^t=-E$ \\ \hline
$Sp(2n)$ & Tr$(E)=0$, & Tr$(E)=0$,\quad or\quad Tr$(E)=\frac{2ih}{c\pi}$,
\\
& $JE^tJ=\pm E$ & $JE^tJ=E$\qquad\quad$JE^tJ=-E$ \\ \hline
$SU(N)$- & $E^t=\pm E$ & no solution \\
conjugating & & \\ \hline
\end{tabular}
\end{center}
\vspace{1ex}

\subsubsection{$SU(N)$}

We begin by considering $SU(N)$, where in addition to the
constraints imposed by unitarity and analyticity we have a single
extra constraint on the trace of $E$. From the first two
constraints we can express $E$ as the conjugate of a diagonal
matrix $X$ by an $SU(N)$ matrix, $$E=Q^{\dag}XQ\qquad{\rm with}\
\; Q\in SU(N)$$ where $X$ is of the form $$X=\left(
\begin{array}{cc}
I_M & 0 \\
0 & -I_{N-M}
\end{array}\right)\,,$$
$I_M$ is the $M \times M$ identity matrix and $0\leq M\leq N$ (see
appendix ~\ref{ssubsec:sug}). By the cyclicity of trace, if we
impose the trace condition for $K_1$ then $N$ must be even and
equal to $2M$. If we impose the condition for $K_2$ we obtain a
condition on $c$ and $d$, and so find
\begin{center}
\begin{tabular}{|c|c|}\hline &\\[-0.15in]
$K_1(\theta)$ & $K_2(\theta)$ \\ \hline &\\[-0.15in]
$E=Q_1^{\dag}\left(
\begin{array}{cc}
I_{N/2} & 0 \\
0 & -I_{N/2}
\end{array}\right)Q_1$ &
$E=Q_2^{\dag}\left(
\begin{array}{cc}
I_M & 0 \\
0 & -I_{N-M}
\end{array}\right)Q_2$\phantom{M} {\small where}\ $\frac{1}{c}+
\frac{1}{d}=\frac{\pi(2M-N)}{ih}$\\
\hline
\end{tabular}
\end{center}
\vspace{1ex} with $Q_i \in SU(N)$. We can see that the case $K_1$
corresponds to the limit of $K_2$ in which we take
$M=\frac{N}{2}$, that is the $c,d\to\infty$, as we would expect.

Thus we have parametrized the possibilities for $E$ with a matrix
$Q\in SU(N)$ and an integer $M$. Once $M$ is fixed, the suitable
$E$ form a space isomorphic to the symmetric space
$$\frac{SU(N)}{S(U(M)\times U(N-M))}$$ where the correspondence is
between an element $E=Q^{\dag}XQ$ and the left coset of $H=S(U(M)\times
U(N-M))$ by $Q$. In the same way the possible $E$ for $K_1(\theta)$ form
a space isomorphic to
$$\frac{SU(N)}{S(U(N/2)\times U(N/2))}\qquad {\rm (only \;for}\ N\
{\rm even}).$$

\subsubsection{$SO(N)$}
\label{ssubsec:ceso}

We now consider $SO(N)$, where in addition to the constraints
associated with $SU(N)$ we also have $E^t=\pm E$. We consider
first the case $E^t=E$: $$E^{\dag}=E\,,\ E^t=E\;\Rightarrow\;
E^{\ast}=E.$$ So $E$ is a symmetric real matrix, and we can
diagonalize it by conjugating with a matrix $R\in SO(N)$. Since
$E$ squares to the identity the diagonal matrix must be of the
form $$X=\left(
\begin{array}{cc}
I_M & 0 \\
0 & -I_{N-M}
\end{array}\right).$$
Then imposing the constraints on the trace of $E$ is the same as for
$SU(N)$ and we have
\begin{center}
\begin{tabular}{|c|c|}\hline &\\[-0.15in]
$K_1(\theta)$ & $K_2(\theta)$ \\ \hline &\\[-0.15in]$E=R_1^t\left(
\begin{array}{cc}
I_{N/2} & 0 \\
0 & -I_{N/2}
\end{array}\right)R_1$ &
$E=R_2^t\left(
\begin{array}{cc}
I_M & 0 \\
0 & -I_{N-M}
\end{array}\right)R_2$\phantom{M} {\small where}\ $c=\frac{2ih}{\pi(2M-N)}$\\
\hline
\end{tabular}
\end{center}
\vspace{1ex} with $R_i \in SO(N)$. In a similar way to the $SU(N)$
case, once $M$ is fixed, the space of matrices $E$ is isomorphic
to $$\frac{SO(N)}{S(O(M)\times O(N-M))}\,,$$ with the $E$ for
$K_1(\theta)$ isomorphic to
$$\frac{SO(N)}{S(O(N/2)\times O(N/2))}\,.$$ Thus, in the same way
as for $SU(N)$, we have an isomorphism between the space of
allowed $E$ and the symmetric spaces (see
appendix~\ref{ssubsec:sog} for more details).

The remaining case to consider is that of antisymmetric $E$. We
find in this case (see appendix~\ref{ssubsec:sou}) that the
matrices $E$ form a space isomorphic to two copies of the
symmetric space $$\frac{SO(N)}{U(N/2)}\,.$$

\subsubsection{$S\!p(N)$}

In addition to the $SU(N)$ constraints, for $S\!p(N)$ we also have
$JE^tJ=\pm E$. We consider first the case $JE^tJ=-E$, with
$$E^2=I\quad{\rm and}\quad JE^tJ=-E\Rightarrow
JE^tJE=-I\Rightarrow E^tJE=J$$ so that, since we also know $E\in
U(N)$, we must have $E\in S\!p(N)$. After appealing to an argument
involving quarternionic matrices (appendix~\ref{ssubsec:spg}) we
find that the space of allowed $E$ for $K_2$ is isomorphic to
$$\frac{S\!p(N)}{S\!p(M)\times S\!p(N-M)}\,.$$ In the case of
$K_1(\theta)$ we again require $M=\frac{N}{2}$ and the $E$-space
is isomorphic to $$\frac{S\!p(N)}{S\!p(N/2)\times S\!p(N/2)}\,.$$

For $K_1(\theta)$ with $JE^tJ=+E$,  $E$ is conjugate over
$\mathbb{C}$ to $$\left(\begin{array}{cc} I_{N/2} & 0 \\ 0 &
-I_{N/2}
\end{array}\right)$$
as $Tr(E)=0$. We find (see appendix~\ref{ssubsec:spu}) that the
allowed $E$ form a space isomorphic to $$\frac{S\!p(N)}{U(N/2)}.$$

\subsubsection{$SU(N)$-conjugating}
\label{ssubsec:cesuc}

The last case to consider is that of $SU(N)$-conjugating. We recall
that in this case the constraints  due to unitarity and
analyticity were slightly modified, to $$E^{\dag}E=I\quad{\rm
and}\quad \det{E}=1\,.$$ Crossing-unitarity imposed $E^t=\pm
E$. Taking the symmetric case first, the allowed $E$ form a set
$$\{E|E^{\dag}E=I, E^t=E, \det{E}=1\}$$ which turns out (see
appendix~\ref{ssubsec:suso}) to be isomorphic to
$$\frac{SU(N)}{SO(N)}.$$

Lastly, we turn to the antisymmetric case $$\{E|E^{\dag}E=I,
E^t=-E, \det{E}=1\}\,.$$ This we find
(appendix~\ref{ssubsec:susp}) is isomorphic to
$$\{1,\omega^2\}\times\frac{SU(N)}{S\!p(N)}\qquad{\rm
where\ any}\ \omega\ {\rm s.t.}\ \omega^N=-1\ {\rm is\ chosen.}$$

\section{The PCM boundary $S$-matrices}
\label{sec:pcmbsm}

In this section we construct the boundary $S$-matrices for the
principal chiral model. We recall~\cite{ogiev87} that the bulk
model $S$-matrix has $G\times G$ symmetry and is constructed as
\be
S_{PCM}(\theta)=X_{11}(\theta)\Big(S_L(\theta)\otimes
S_R(\theta)\Big) \ee where $X_{11}(\theta)$ is the CDD factor for
the PCM and $S_{L,R}(\theta)$ are left and right copies of the
minimal $S$-matrix possessing $G$-symmetry. Following this
prescription, we shall use the minimal $K$-matrices from the
previous section to construct boundary $S$-matrices for the PCM on
a half-line. We then go on to explore their symmetries and make
connection with the classical results of section two.

\subsection{The CDD factors}
\label{subsec:cddfac}

Introducing the CDD factor, $X_{11}(\theta)$, into the bulk
$S$-matrix for the PCM requires that we introduce an extra factor,
$Y_{11}(\theta)$ (or $Y_{1\bar{1}}(\theta)$ in the case of $SU(N)$
conjugating), into the boundary $S$-matrix in order to satisfy
crossing-unitarity. We construct the boundary $S$-matrix for the
PCM as
\be
K_{PCM}(\theta)=Y_{11}(\theta)\Big(K_L(\theta)\otimes
K_R(\theta)\Big) \ee where $K_{L,R}(\theta)$ are left and right
copies of the same type of minimal $K$-matrix, chosen from among
the possibilities classified in section three. In order that
$K_{PCM}(\theta)$ satisfy the crossing-unitarity equation with
$S_{PCM}(\theta)$ we require
\bea
Y_{11}(i\pi-\theta)Y_{11}(i\pi+\theta)=1\quad&\quad SU(N), \nonumber \\
\frac{Y_{11}(\frac{i\pi}{2}-\theta)}{Y_{11}(\frac{i\pi}{2}+\theta)}=
X_{11}(2\theta)\quad&\quad SO(N)\ {\rm and}\ S\!p(N), \\
\frac{Y_{1\bar{1}}(\frac{i\pi}{2}-\theta)}{Y_{1\bar{1}}(\frac{i\pi}{2}+
\theta)}=X_{11}(2\theta)\quad&\quad SU(N)-{\rm conjugating}. \nonumber \eea

The CDD factors for the bulk PCM $S$-matrices are\bea
X_{11}(\theta)=(2)_{\theta}=X_{1\bar{1}}(i\pi-\theta)\quad&\quad
SU(N), \nonumber \\
X_{11}(\theta)=(2)_{\theta}(h-2)_{\theta}\qquad\quad\,&\quad
SO(N)\ {\rm and}\ S\!p(N), \eea
$${\rm where}\quad(x)_{\theta}=\frac{\sinh{(\frac{\theta}{2}+\frac{i\pi
x}{2h})}}{\sinh{(\frac{\theta}{2}-\frac{i\pi x}{2h})}}\,.$$ We
find~\cite{fring93} the following candidates for the $Y$
functions: \bea
Y_{11}(\theta)=-(1-h)_{\theta}\quad\qquad\qquad\qquad\qquad&SU(N)\qquad
 \nonumber \\
Y_{11}(\theta)=-\left(\frac{h}{2}+2\right)_{\theta}\left(\frac{h}
{2}+1\right)_{\theta}(1-h)_{\theta}&SO(N)\ {\rm and}\ S\!p(N) \\
Y_{1\bar{1}}(\theta)=\left(\frac{h}{2}+2\right)_{\theta}\left(\frac{h}
{2}+1\right)_{\theta}\ \ \quad\qquad&\qquad SU(N)-{\rm
conjugating}, \nonumber \eea and note that none of these factors
have poles on the physical strip.

We still have freedom to multiply by an arbitrary boundary CDD
factor. That is, we can replace any of the above $Y(\theta)$
factors by $g(\theta)Y(\theta)$, where $g(\theta)$ is a CDD
factor. This allows us to introduce simple poles into the PCM
boundary $S$-matrices. In the case where we construct a PCM
$S$-matrix using left and right copies of $K_2(\theta)$, we wish
to introduce the CDD factor\footnote{We are assuming, as in
section~\ref{subsec:bsm}, that $M\leq\frac{N}{2}$ so that $
\frac{1}{c}$ is on the physical strip. As stated in
section~\ref{subsec:bsm}, the resulting PCM boundary $S$-matrix
will possess a $c\leftrightarrow -c$ (and $d\leftrightarrow -d$ for
$G\!=\!SU(N)$) symmetry and so will be correct for $M\geq\frac{N}{2}$ also.}
\bea
g(\theta)=\left(\frac{h}{ci\pi}\right)_{\theta}\left(h-\frac{h}{di\pi}
\right)_{\theta} & SU(N), \nonumber \\
g(\theta)=\left(\frac{h}{ci\pi}\right)_{\theta}\left(h-\frac{h}{ci\pi}
\right)_{\theta} & SO(N)\ {\rm and}\ S\!p(N) \eea which gives a simple pole at
$\theta=\frac{1}{c}$, corresponding to the formation of a boundary
bound state.

\subsection{The boundary $S$-matrices}
\label{subsec:tbsm}

We now list the full PCM boundary $S$-matrices for the various
$G$. We make use of the relation
$$\Gamma(z)\Gamma(1-z)=\frac{\pi}{\sin{(\pi z)}}\,.$$ We will also
require the scalar factors
$$\eta(\theta)=\frac{\Gamma\left(\frac{\theta}{2i\pi}+\frac{1}{2}+
\frac{1}{2h}\right)\Gamma\left(\frac{-\theta}{2i\pi}+\frac{1}{2}\right)
}{\Gamma\left(\frac{-\theta}{2i\pi}+\frac{1}{2}+\frac{1}{2h}\right)
\Gamma\left(\frac{\theta}{2i\pi}+\frac{1}{2}\right)}\,
,\qquad\nu(\theta
)=\frac{\Gamma\left(\frac{\theta}{2i\pi}+\frac{1}{2}+\frac{1}{2h}\right)
\Gamma\left(\frac{-\theta}{2i\pi}\right)}{\Gamma\left(\frac{-\theta}{
2i\pi}+\frac{1}{2}+\frac{1}{2h}\right)\Gamma\left(\frac{\theta}{2i\pi}
\right)}\, ,$$
$$\mu(\theta)=\frac{1}{4\pi^2c^2}\frac{\Gamma\left(\frac{-\theta}{2i\pi
}+\frac{1}{2}+\frac{1}{2i\pi
c}\right)\Gamma\left(\frac{-\theta}{2i\pi}+
\frac{1}{2}-\frac{1}{2i\pi
c}\right)\Gamma\left(\frac{\theta}{2i\pi}+ \frac{1}{2i\pi
c}\right)\Gamma\left(\frac{\theta}{2i\pi}-\frac{1}{2i\pi
c}\right)}{\Gamma\left(\frac{\theta}{2i\pi}+\frac{1}{2}+\frac{1}{2i\pi
c
}\right)\Gamma\left(\frac{\theta}{2i\pi}+\frac{1}{2}-\frac{1}{2i\pi
c} \right)\Gamma\left(\frac{-\theta}{2i\pi}+1+\frac{1}{2i\pi
c}\right) \Gamma\left(\frac{-\theta}{2i\pi}+1-\frac{1}{2i\pi
c}\right)}\,,$$
$$\lambda(\theta)=\frac{1}{4\pi^2c^2}\frac{\Gamma\left(\frac{-\theta}{2i
\pi}+\frac{M}{N}+\frac{1}{2i\pi c}\right)\Gamma\left(\frac{-\theta}{2i
\pi}+1-\frac{M}{N}-\frac{1}{2i\pi c}\right)\Gamma\left(\frac{\theta}{2i
\pi}+\frac{1}{2i\pi c}\right)\Gamma\left(\frac{\theta}{2i\pi}-\frac{1}{2
i\pi c}\right)}{\Gamma\left(\frac{\theta}{2i\pi}+\frac{M}{N}+\frac{1}{2i
\pi c}\right)\Gamma\left(\frac{\theta}{2i\pi}+1-\frac{M}{N}-\frac{1}{2i
\pi c}\right)\Gamma\left(\frac{-\theta}{2i\pi}+1+\frac{1}{2i\pi c}
\right)\Gamma\left(\frac{-\theta}{2i\pi}+1-\frac{1}{2i\pi c}\right)}$$
$${\rm and}\qquad\epsilon_{n,m}(\theta)=\frac{\Gamma\left(\frac{\theta}
{2i\pi}+\frac{n}{4}\right)\Gamma\left(\frac{-\theta}{2i\pi}+\frac{m}{4}
+\frac{1}{2h}\right)}{\Gamma\left(\frac{-\theta}{2i\pi}+\frac{n}{4}
\right)\Gamma\left(\frac{\theta}{2i\pi}+\frac{m}{4}+\frac{1}{2h}\right)}
\,.$$
The PCM $S$-matrices can then be written as follows.

\subsubsection{$SU(N)$}

We have found two types of boundary $S$-matrix for $SU(N)$,
\be
K_{PCM}(\theta)=-(1-h)_{\theta}\,\mu(\theta)\Big(\nu(\theta)(I+c\theta
E_L) \otimes\nu(\theta)(I+c\theta E_R)\Big)
\ee
(whose $c\to\infty$ limit:
\be
K_{PCM}(\theta)=-(1-h)_{\theta}\Big(\eta(\theta)E_L\otimes\eta(\theta)
E_R\Big) \ee
is a valid PCM boundary scattering matrix) where\footnote{We are using
the symmetric space notation $G/H$ here to denote the relevant translated
Cartan immersion. Details are given in appendix~\ref{sec:apssci}.}
$$E_{L/R}\in\frac{SU(N)}{S(U(N/2)\times U(N/2))}\,,$$ and
\be\label{pcmsu}
K_{PCM}(\theta)=-(1-h)_{\theta}\,\lambda(\theta)\Big(\nu(\theta)(I+c\theta
E_L) \otimes\nu(\theta)(I+c\theta E_R)\Big) \ee where
$$E_{L/R}\in\frac{SU(N)}{S(U(N-M)\times U(M))}\,.$$

\subsubsection{$SO(N)$}

For $SO(N)$ three types have been found,
\be
K_{PCM}(\theta)=-\left(\frac{h}{2}+2\right)_{\theta}\left(\frac{h}{2}
+1\right)_{\theta}(1-h)_{\theta}\mu(\theta)\Big(\nu(\theta)
\epsilon_{3,3}(\theta)(I+c\theta E_L)\otimes\nu(\theta)\epsilon_{3,3}(
\theta)(I+c\theta E_R)\Big)
\ee
(whose $c\to\infty$ limit:
\be
K_{PCM}(\theta)=-\left(\frac{h}{2}+2\right)_{\theta}\left(\frac{h}{2}
+1\right)_{\theta}(1-h)_{\theta}\Big(\eta(\theta)\epsilon_{3,3}(\theta)
E_L\otimes\eta(\theta)\epsilon_{3,3}(\theta)E_R\Big) \ee
is a valid PCM boundary scattering matrix) where\footnote{The factor
$\{+1,-1\}$ indicates that the space containing $E_{L/R}$ is a twofold
copy of the symmetric space -- no group structure is implied. See
appendix~\ref{ssubsec:sou} for details.}
$$E_{L/R}\in\frac{SO(N)}{U(N/2)}\times\{+1,-1\}\,,$$

\be
K_{PCM}(\theta)=-\left(\frac{h}{2}+2\right)_{\theta}\left(\frac{h}{2}
+1\right)_{\theta}(1-h)_{\theta}\Big(\eta(\theta)\epsilon_{3,1}(\theta)
E_L\otimes\eta(\theta)\epsilon_{3,1}(\theta)E_R\Big) \ee where
$$E_{L/R}\in\frac{SO(N)}{S(O(N/2)\times O(N/2))}\,,$$ and
\be\label{pcmso}
K_{PCM}(\theta)=-\left(\frac{h}{2}+2\right)_{\theta}\left(\frac{h}{2}
+1\right)_{\theta}(1-h)_{\theta}\mu(\theta)\Big(\nu(\theta)
\epsilon_{3,1}(\theta)(I+c\theta
E_L)\otimes\nu(\theta)\epsilon_{3,1}( \theta)(I+c\theta E_R)\Big)
\ee where $$E_{L/R}\in\frac{SO(N)}{S(O(N-M)\times O(M))}\qquad{\rm and}\
c=\frac{2ih}{\pi(2M-N)}\ {\rm in}\ \mu(\theta).$$

\subsubsection{$S\!p(N)$}

Three types of $K_{PCM}(\theta)$ have also been found for
$S\!p(N)$,
\be
K_{PCM}(\theta)=-\left(\frac{h}{2}+2\right)_{\theta}\left(\frac{h}{2}
+1\right)_{\theta}(1-h)_{\theta}\mu(\theta)\Big(\nu(\theta)
\epsilon_{3,1}(\theta)(I+c\theta E_L)\otimes\nu(\theta)\epsilon_{3,1}(
\theta)(I+c\theta E_R)\Big)
\ee
(whose $c\to\infty$ limit:
\be
K_{PCM}(\theta)=-\left(\frac{h}{2}+2\right)_{\theta}\left(\frac{h}{2}
+1\right)_{\theta}(1-h)_{\theta}\Big(\eta(\theta)\epsilon_{3,1}(\theta)
E_L\otimes\eta(\theta)\epsilon_{3,1}(\theta)E_R\Big) \ee
is a valid PCM boundary scattering matrix) where
$$E_{L/R}\in\frac{S\!p(N)}{U(N/2)}\,,$$

\be
K_{PCM}(\theta)=-\left(\frac{h}{2}+2\right)_{\theta}\left(\frac{h}{2}
+1\right)_{\theta}(1-h)_{\theta}\Big(\eta(\theta)\epsilon_{3,3}(\theta)
E_L\otimes\eta(\theta)\epsilon_{3,3}(\theta)E_R\Big) \ee where
$$E_{L/R}\in\frac{S\!p(N)}{S\!p(N/2)\times S\!p(N/2))}\,,$$ and
\be\label{pcmsp}
K_{PCM}(\theta)=-\left(\frac{h}{2}+2\right)_{\theta}\left(\frac{h}{2}
+1\right)_{\theta}(1-h)_{\theta}\mu(\theta)\Big(\nu(\theta)
\epsilon_{3,3}(\theta)(I+c\theta
E_L)\otimes\nu(\theta)\epsilon_{3,3}( \theta)(I+c\theta E_R)\Big)
\ee where $$E_{L/R}\in\frac{S\!p(N)}{S\!p(N-M)\times
S\!p(M))}\qquad{\rm and}\ c=\frac{2ih}{\pi(2M-N)}\ {\rm in}\ \mu(\theta)
.$$

\subsubsection{$SU(N)$-conjugating}

Lastly, we have found two types of representation-conjugating
boundary $S$-matrix for $SU(N)$
\be
K_{PCM}(\theta)=\left(\frac{h}{2}+2\right)_{\theta}\left(\frac{h}{2}
+1\right)_{\theta}\Big(\epsilon_{1,1}(\theta)E_L\otimes\epsilon_{1,1}(
\theta)E_R\Big) \ee where
$$E_{L/R}\in\frac{SU(N)}{SO(N)}\,,$$
and
\be
K_{PCM}(\theta)=\left(\frac{h}{2}+2\right)_{\theta}\left(\frac{h}{2}
+1\right)_{\theta}\Big(\epsilon_{1,3}(\theta)E_L\otimes\epsilon_{1,3}(
\theta)E_R\Big) \ee where
$$E_{L/R}\in\{1,\omega^2\}\times\frac{SU(N)}{S\!p(N)}\qquad(\omega^N=-1)
\,.$$

\subsection{Symmetries of the PCM boundary $S$-matrices}

We now consider the symmetries possessed by the PCM boundary
$S$-matrices. Before looking at the surviving group symmetries at
the boundary, we first point out a symmetry possessed by those
$S$-matrices constructed from the $K_2(\theta)$-type minimal
solution.

\subsubsection{$M\leftrightarrow N-M$ symmetry}

This was first noted, for $SU(N)$ diagonal boundary scattering, in
\cite{doikou98}. The Grassmannian symmetric spaces
$$\frac{SU(N)}{S(U(M)\times
U(N-M))}\qquad\frac{SO(N)}{S(O(M)\times
O(N-M))}\qquad\frac{S\!p(N)}{S\!p(M)\times S\!p(N-M)}$$ are all
invariant under $M\leftrightarrow N-M$. Consequently, we might
expect that the PCM boundary $S$-matrices constructed using
matrices $E_{L/R}$ lying in translated Cartan constructions of
these symmetric spaces would also respect this symmetry. This is
exactly what we find for the $K_{PCM}(\theta)$
matrices~(\ref{pcmsu}),~(\ref{pcmso}) and~(\ref{pcmsp}).

To see this invariance, we consider how the exchange $M\leftrightarrow
N-M$ affects the degrees of freedom in the $K$-matrices. The matrices
$E_{L/R}$ are constructed as
\be
E_{L/R}=U_{L/R}XU_{L/R}^{-1}\qquad{\rm where}\quad U_{L/R}\in
SU(N), SO(N)\ {\rm or}\ S\!p(N)\quad {\rm and} \ee
$$X=\left(\begin{array}{c|c} I_M & 0
\\ \hline 0 & -I_{N-M}
\end{array}\right),\qquad{\rm so\ under}\ M\leftrightarrow N-M\qquad
X\mapsto\check{X}=\left(\begin{array}{c|c} I_{N-M} & 0 \\ \hline 0
& -I_M
\end{array}\right).$$
Thus under $M\leftrightarrow N-M,\;
E_{L/R}\mapsto\check{E}_{L/R}=U_{L/R}\check{X}U_{L/R}^{-1}$.
Taking traces we see that $c\leftrightarrow -c$ (and $d\leftrightarrow
-d$) under the exchange. Now note that the scalar factor $\mu(\theta)$
is invariant under $c\leftrightarrow -c$ and $\lambda(\theta)$ is
invariant under $c\leftrightarrow -c$, $M\leftrightarrow N-M$. So
$K_{PCM}(\theta)\mapsto\check{K}_{PCM}(\theta)$ where $$(I+c\theta
E_{L/R})\mapsto(I-c\theta\check{E}_{L/R})\,.$$ Now
$$-\check{E}_{L/R}=U_{L/R}(-\check{X})U_{L/R}^{-1}=U_{L/R}OXO^{-1}U_{L/R
}^{-1}$$ where $$O=\left(\begin{array}{c|c} 0 & I_{N-M} \\ \hline
\pm I_M & 0
\end{array}\right).$$
We choose the sign $\pm$ to ensure that $\det{O}=1$ and (noting
that when $N$ and $M$ are even $O\in S\!p(N)$) we see that
$U_{L/R}\in G=SU(N)$, $SO(N)$ or $S\!p(N)\ \Longrightarrow
(U_{L/R}O)\in G$.

Thus, if we denote by $K_{PCM}(\theta;U_L, U_R)$ the matrix
constructed using \be\label{KLR} E_{L/R}=U_{L/R}XU_{L/R}^{-1}\ee
and by $\check{K}_{PCM}(\theta;U_L, U_R)$ the image of this under
the exchange $M\leftrightarrow N-M$, we have
\be
\check{K}_{PCM}(\theta;U_L,U_R)=K_{PCM}(\theta;U_LO,U_RO)\,. \ee
So we see that the $K_{PCM}(\theta)$ matrices do respect this
invariance of the symmetric spaces, in the sense that the action
of the exchange $M\leftrightarrow N-M$ on the $K$-matrices is
simply a translation in the parameter space.

A further consequence of this emerges if we consider the pole
structure of these $K$-matrices. Restricting to the choice of parameter
$c=d=\frac{2iN}{\pi(2M-N)}$ in the case $G=SU(N)$, so that it is
analogous to the $G=SO(N), S\!p(N)$ cases, there is exactly one simple
pole on the physical strip at either $\theta=\frac{1}{c}$ or $\theta=-
\frac{1}{c}$ (since $M\neq\frac{N}{2}$). If we interpret the simple pole
as the formation of a boundary bound state at this rapidity, then the
bound state is in a representation projected onto by either
$P_L^+\otimes P_R^+$ or $P_L^- \otimes P_R^-$, respectively, where
\be
P_{L/R}^{\pm}=\frac{1}{2}(I\pm E_{L/R})=U_{L/R}\left(\frac{1}{2}(I\pm
X)\right)U_{L/R}^{-1}
\ee
We note
$$\frac{1}{2}(I+X)=\left(\begin{array}{c|c}
I_M & 0 \\ \hline
0 & 0
\end{array}\right)\qquad{\rm
and}\qquad\frac{1}{2}(I-X)=\left(\begin{array}{c|c}
0 & 0 \\ \hline
0 & I_{N-M}
\end{array}\right)\,.$$
We find that $\pm\frac{1}{c}$ lies on the physical strip as
$M\lessgtr\frac{N}{2}$, and so the boundary bound state
representation is always the smaller of the two projection spaces.
We plan to investigate further the spectrum of boundary bound
states in future work, but for the moment we return to consider
the surviving remnant of group symmetry at the boundary and make
connections with the classical boundary conditions of section two.

\subsubsection{Boundary group symmetry in the non-conjugating cases}

We recall~\cite{ogiev87} that the principal chiral model in the
bulk possesses a global $G\times G$ symmetry, respected by the
bulk $S$-matrices. In section two we saw that the introduction of
a boundary in the classical PCM generally breaks the $G\times G$
symmetry, so that only a remnant survives, the nature of which is
dictated by the boundary condition. In particular we saw in
section~\ref{ssubsec:phlcbc} that the boundary
condition~(\ref{bc2}), $$g(0)\in k_LH k_R^{-1}\qquad{\rm where}\
k_{L/R}\ {\rm parametrize\ left/right\ cosets\ of}\ H\in G,$$
preserves $k_LHk_L^{-1}\times k_RHk_R^{-1}$.

Turning our attention to the PCM boundary $S$-matrices, we find
that $K_{PCM}(\theta;k_L,k_R)$ is invariant under exactly this
symmetry. That is,
\be
\left[K_{PCM}(\theta;k_L,k_R),k_LHk_L^{-1}\times
k_RHk_R^{-1}\right]=0\,. \ee We begin with the Grassmannian cases,
where it is enough to show that
$$\left[(I+c\theta E_L)\otimes(I+c\theta E_R),k_Lh_Lk_L^{-1}\times
k_Rh_Rk_R^{-1}\right]=0\,,$$ where (for subscripts $L$ and $R$)
$h$ are arbitrary elements of $H$, $E=kXk^{-1}$ and $k\in
G=SU(N)$, $SO(N)$ or $S\!p(N)$.
But this is immediate: $Xh=hX$, since $H$ is constructed to be
precisely those elements in $G$ which commute with $X$.

For the case $G/H=SO(2n)/U(n)$ (respectively $G/H=S\!p(2n)/U(n)$)
we note (appendix \ref{ssubsec:sou}, resp.\ \ref{ssubsec:spu})
that $E=ikJk^{-1}$ where $k\in G$, and that $H=U(n)$ is constructed
as those elements in $G$ satisfying $Jh=hJ$, giving the required
result.

\subsubsection{Boundary group symmetry in the $SU(N)$-conjugating case}

The cases of $SU(N)/SO(N)$ and $SU(N)/S\!p(N)$ are a little more
subtle.  Performing similar calculations to the above (and again
leaving the $L/R$ suffix implicit) we find, on constructing $E$ as
in appendices \ref{ssubsec:suso}, \ref{ssubsec:susp}, that
$E\,khk^{-1}  =(khk^{-1})^{\ast}E$, which implies
\be
 K_{PCM}(\theta;k_L,k_R)\,(k_LH_Lk_L^{-1}\times
k_RH_Rk_R^{-1})=(k_LH_Lk_L^{-1}\times
k_RH_Rk_R^{-1})^{\ast}\,K_{PCM} (\theta;k_L,k_R)\,. \ee Such a
result is not surprising, since in this case
$K_{PCM}(\theta):V_L\otimes V_R\to\bar{V}_L\otimes\bar{V}_R$. It
is straightforward to obtain a symmetry relation in which, as
earlier in section~\ref{ssubsec:uhac}, we consider a boundary
$S$-matrix which is an endomorphism of
$(V_L\oplus\bar{V}_L)\otimes(V_R\oplus\bar{V}_R)$. We do not give
details.

\subsection{Concluding summary}

When $G/H$ is a symmetric space, the classical boundary condition
$g(0)\in k_LH k_R^{-1}$ preserves the local PCM conserved charges
necessary for integrability. Thus, as stated in section
~\ref{ssubsec:phlcbc}, the possible BCs are parametrized by a
moduli space $G/H\times G/H$\footnote{up to a discrete ambiguity,
as further explained in appendix~\ref{subsec:acbcs}}. We have also
found boundary $S$-matrices which are parametrized by $G/H\times
G/H$. Further, we find that the global symmetry which survives in
the presence of this BC is precisely that which commutes with the
boundary $S$-matrix $K_{PCM}(\theta;k_L,k_R)$\footnote{with the
subtlety noted above in the case of $SU(N)$-conjugated}. So we
finish with this \\
{\bf Claim:} The principal chiral model on $G$ is
classically integrable with boundary condition $g(0)\in k_L H
k_R^{-1}$, where $k_{L/R}\in G$ and $G/H$ is a symmetric space;
and it remains integrable at the quantum level, where its boundary
$S$-matrix is $K_{PCM}(\theta;k_L,k_R)$.

\vspace{0.1in} \noindent {\bf Additional comment for v3}

Recently a paper~\cite{arnau03} has appeared which deals with
$osp(m|n)$ spin chain models. Its results are related to ours for
the $G=SO(N)$ and $G=S\!p(N)$ cases.

\vspace{0.1in} \noindent {\bf Acknowledgments}

We should like to thank Tony Sudbery and Ian McIntosh for
discussions of symmetric spaces. Our thanks also go to G\'erard Watts
for pointing out an error in our original discussion of crossing
unitarity. NJM would like to thank Patrick Dorey and Ed Corrigan for
helpful discussions, and Bernard Piette, Paul Fendley and Jonathan Evans
for email exchanges. BJS would like to thank Gustav Delius, Brett Gibson
and Mark Kambites for discussions. Finally NJM thanks the Centre de
Recherches Math\'ematiques, U de Montr\'eal, where this work was
begun during the `Quantum Integrability 2000' program, for
hospitality and financial support, and BJS thanks the UK EPSRC for
a D.Phil.\ studentship.

Finally, we should like to thank G\'erard Watts for pointing out
an error (in our use of the crossing-unitarity relation) in
earlier versions which has led to the changes (namely the slightly
greater freedom in some scalar prefactors) in v3.

\section{Appendix: Symmetric spaces and the Cartan immersion}
\label{sec:apssci}

Under the action of an involutive automorphism $\alpha$ (which may
or may not be inner), a Lie algebra splits into eigenspaces
$\g=\h\oplus\kk$ of eigenvalue $+1$ ($\h$) and $-1$ ($\kk$), with
$$ [\h,\h]\subset\h\;,\qquad [\h,\kk]\subset \kk \;,\qquad
[\kk,\kk]\subset \h\,.$$ The subgroup $H$ generated by $\h$ is
compact, and we have taken $G$ to be compact (type I) rather than
maximally non-compact (type III). For the classical groups these
are the groups $G=SU(N),SO(N)$ and $S\!p(N)$ (where the argument
of $S\!p$ is understood always to be even) themselves, along with
those described in the table below. The dimension is
dim$G-$dim$H$, and the automorphism is given by its action on $U$
in the defining representation, where $X$ is the diagonal matrix
with $M$ $+1$s and $N-M$ $-1$s and $J$ is the symplectic form
matrix, which is block-diagonal with $N/2$ blocks
{\scriptsize${\left(\begin{array}{rr}0&\!-1\\1&0\end{array}\right)}$}
and satisfies $J^2=-I_N$.
\[\begin{array}{|c|c|c|}\hline{\rm symmetric \;\, space} & {\rm
dimension} & {\rm automorphism}\\ \hline & &\\ \hspace{0.2in}

{SU(N)/ S(U(N-M)\times U(M))}\hspace{0.2in} & 2M(N-M)& U\mapsto
XUX
\\[0.1in]

{SO(N)/ SO(N-M)\times SO(M)} & M(N-M) & U\mapsto XUX\\[0.1in]

{S\!p(N) / S\!p(N-M)\times S\!p(M)} & M(N-M) & U\mapsto
XUX\\[0.1in]

{SU(N)/ SO(N)} & {N(N+1)\over 2}-1 & U\mapsto U^{\ast}\\[0.1in]

{SU(N)/ S\!p(N)} & {N(N-1)\over 2}-1 & U\mapsto
-JU^{\ast}J\\[0.1in]

{SO(2n)/ U(n)} &n(n-1)& U\mapsto -JUJ \\[0.1in]

{S\!p(2n)/ U(n)}& n(n+1)& U\mapsto -JUJ\\[0.1in]\hline
\end{array}\]
\centerline{(We refer to the first three as the `Grassmannian'
cases.)}

\subsection{The Cartan immersion}
\label{subsec:tci}

The Cartan immersion constructs $G/H$ as a subspace of $G$ (due to
Cartan, and described briefly in~\cite{helga78} or more fully in
\cite{fomen90} (vol.II, sect.10, prop.4). Lifting $\alpha$ in the
natural way from the algebra to the group (so that $\alpha(h)=h$
for all $h\in H$), under  \begin{eqnarray*} gH & \mapsto &
\alpha(g) g^{-1} \\ {\rm we\;\;have}\qquad G/H & = & \{\alpha(g)
g^{-1}|g\in G\}\,.\end{eqnarray*} (This statement, of course,
depends crucially on the fact that we have chosen $H$ so that it
consists of {\em all} elements of $G$ invariant under $\alpha$;
for the more general case see \cite{EF,AP}.)

We then have $\alpha(k)=k^{-1}$ for all $k\in {G\over
H}\hookrightarrow G$. Defining \be\label{EFman} {\cal M} = \{k\in
G| \alpha(k)=k^{-1}\}\,,\ee it turns out \cite{EF,AP} that $G/H$
is in 1-1 correspondence with ${\cal M}_0$, the identity-connected
component of ${\cal M}$. In the non-Grassmannian cases ${\cal
M}={\cal M}_0$, but in the Grassmannian cases ${\cal M}$ is a
union of disconnected components, each of which is the Cartan
immersion of a different $G/H'$ \cite{AP}.

In order to make connections with
subsection~\ref{subsec:constraints} we consider translations of
 the Cartan-immersed $G/H$. In the
Grassmannian cases we translate by left-multiplying by the
diagonal matrix $X$. (In the unitary and orthogonal cases if
$\det{X}=-1$ then the resulting construction is no longer a subset
of $SU(N)$ or $SO(N)$, but lies in the determinant $-1$ part of
$U(N)$ or $O(N)$, respectively.) In the case of $SU(N)/SO(N)$ we
do not need to translate the Cartan construction. For
$SU(N)/S\!p(N)$ we translate by $J$ (which has determinant $1$).
In the remaining cases of $SO(2n)/U(n)$ and $S\!p(2n)/U(n)$ we
translate by $iJ$, which has determinant $(-1)^n$ and so will be a
translation into the determinant $-1$ part of $O(2n)$ if and only
if $n$ is odd. The full set of translated Cartan immersions is
then \beaa \frac{SU(N)}{S(U(M)\times U(N-M))}&\cong&\{
UXU^{\dag}|U\in SU(N)\} \\ \frac{SO(N)}{S(O(M)\times
O(N-M))}&\cong&\{ UXU^t|U\in SO(N)\} \\
\frac{S\!p(N)}{S\!p(M)\times S\!p(N-M)}&\cong&\{ UXU^{-1}|U\in
S\!p(N)\} \\ \frac{SU(N)}{SO(N)}&\cong&\{U^{\ast}U^{\dag}|U\in
SU(N)\} \\ \frac{SU(N)}{S\!p(N)}&\cong&\{U^{\ast}JU^{\dag}|U\in
SU(N)\} \\ \frac{SO(2n)}{U(n)}&\cong&\{iUJU^t|U\in SO(2n)\} \\
\frac{S\!p(2n)}{U(n)}&\cong&\{iUJU^{-1}|U\in S\!p(2n)\}\,,  \eeaa
and we treat each of these in turn in appendices 5.2.1 - 5.2.7.

\subsection{The Boundary $S$-matrix Constraints}

The aim of this appendix is to show in every case that the above
constructions of the symmetric spaces can be described in terms of
constraints (those from subsection~\ref{subsec:constraints}) on a
single complex $N\times N$ matrix $E\in Gl(N,\mathbb{C})$.

\subsubsection{$\{ UXU^{\dag}|U\in SU(N)\}=\{E|E^{\dag}=E, E^2=I,
Tr(E)=2M-N\}$}
\label{ssubsec:sug}

$\{ UXU^{\dag}|U\in SU(N)\}\subseteq\{E|E^{\dag}=E, E^2=I,
Tr(E)=2M-N\}$\hspace{1em} is obvious.

Now if $E^{\dag}=E$ then $\exists\ U\in SU(N)$ s.t. $E=UDU^{\dag}$,
where $D$ is diagonal.
$$E^2=I \Rightarrow D^2=I\ {\rm so}\ D\ {\rm has\ diagonal\ entries}\
\pm 1$$
Thus, after possible reordering of the diagonal entries (which we absorb
into $U$)
$$D=\left(\begin{array}{cc}
I_{\tilde{M}} & 0 \\
0 & -I_{N-\tilde{M}}
\end{array}\right)$$
The constraint on $Tr(E)$ implies $D=X$ and so we have $E=UXU^{\dag}$, as
required.

\subsubsection{$\{ UXU^t|U\in SO(N)\}=\{E|E^{\dag}=E, E^2=I, E^t=E,
Tr(E)=2M-N\}$}
\label{ssubsec:sog}

$\{ UXU^t|U\in SO(N)\}\subseteq\{E|E^{\dag}=E, E^2=I, E^t=E,
Tr(E)=2M-N\}$\hspace{1em} is obvious.

Now $E^{\dag}=E$ and $E^t=E$ imply that $E$ is a real symmetric matrix,
therefore
$$\exists\ U\in SO(N)\ {\rm s.t.}\ E=UDU^t,\ {\rm where}\ D\ {\rm
is\ diagonal}$$
As in the case above the conditions $E^2=I$ and $Tr(E)=2M-N$ require
that $D=X$, so we have $E=UXU^t$, as required.

\subsubsection{$\{ UXU^{-1}|U\in S\!p(N)\}=\{E|E^{\dag}=E, E^2=I,
JE^tJ=-E, Tr(E)=2M-N\}$}
\label{ssubsec:spg}

$\{ UXU^{-1}|U\in S\!p(N)\}\subseteq\{E|E^{\dag}=E, E^2=I,
JE^tJ=-E, Tr(E)=2M-N\}$\hspace{1em} is obvious.

In order to show the converse we consider the following
correspondence\footnote{Thanks to Ian McIntosh for providing this
suggestion.}. Consider the representation of $\mathbb{H}$ by the
$2\times 2$ complex matrices
$$\qa=a_0\qe+a_1\qi+a_2\qj+a_3\qk\mapsto
\ma=a_0\me+a_1\mi+a_2\mj+a_3\mk$$ where
$$\me=\left(\begin{array}{cc} 1 & 0 \\ 0 & 1
\end{array}\right)\qquad
\mi=\left(\begin{array}{cc}
i & 0 \\
0 & -i
\end{array}\right)\qquad
\mj=\left(\begin{array}{cc}
0 & -1 \\
1 & 0
\end{array}\right)\qquad
\mk=\left(\begin{array}{cc}
0 & -i \\
-i & 0
\end{array}\right)$$
If we define complex conjugation on $\mathbb{H}$ in the standard way
(note that this corresponds to hermitian conjugation of the matrices),
and consider quaternions of unit length
$$\qa\qa^{\ast}=(a_0^2+a_1^2+a_2^2+a_3^2)\qe=\qe$$
these correspond to elements of $SU(2)=S\!p(2)$ under the map. (Recall
our definition of $S\!p(2n)$, as being the subset of $U(2n)$ satisfying
the condition $AJA^t=J$. In the $n=1$ case this is simply the
constraint $\det{\ma}=1$, and so $SU(2)=S\!p(2)$.)

This correspondence can be generalized to $S\!p(2n)$ in the
following way. Consider $$\mathcal{A}=\left(\begin{array}{cccc}
\qa_{11} & \qa_{12} & \dots & \qa_{1n} \\ \qa_{21} & \qa_{22} &
\dots & \qa_{2n} \\ \vdots & \vdots & \ddots & \vdots \\ \qa_{n1}
& \qa_{n2} & \dots & \qa_{nn}
\end{array}\right)\mapsto
\left(\begin{array}{c|c|c|c}
\ma_{11} & \ma_{12} & \dots & \ma_{1n} \\ \hline
\ma_{21} & \ma_{22} & \dots & \ma_{2n} \\ \hline
\vdots & \vdots & \ddots & \vdots \\ \hline
\ma_{n1} & \ma_{n2} & \dots & \ma_{nn}
\end{array}\right)=A$$
where the quaternion, $\qa_{ij}\mapsto \ma_{ij}$, the $2\times 2$ block, in
the way described above. We define
$$J=\left(\begin{array}{c|c|c|c}
\mj & 0 & \dots & 0 \\ \hline
0 & \mj & & 0 \\ \hline
\vdots & & \ddots & \\ \hline
0 & 0 & & \mj
\end{array}\right)\qquad{\rm where}\;\ \mj=\left(\begin{array}{cc}
0 & -1 \\
1 & 0
\end{array}\right)\,;$$
then the conditions $\mathcal{A}\mathcal{A}^{\dag}=\mathcal{I}$
(where $\mathcal{I}$ is the quaternionic identity matrix) and
$A\in S\!p(2n)$ are in exact correspondence.

We now appeal to the fact that the (quaternionic) unitary matrix
$\mathcal{A}$ can be diagonalized
$$\mathcal{Q}^{\dag}\mathcal{A}\mathcal{Q}=\mathcal{D}\qquad{\rm
where}\;\ \mathcal{Q},\mathcal{D}\in U(n,\mathbb{H})\ \;{\rm and}\
\mathcal{D}\ {\rm is\ diagonal}$$ Under the isomorphism we have
established between $U(n,\mathbb{H})$ and $S\!p(2n)$ this
statement corresponds to $Q^{\dag}AQ=D$ where $Q,D\in S\!p(2n)$
and $D$ is of the form $$D=\left(\begin{array}{c|c|c|c}
\mathbb{D}_1 & 0 & & \\ \hline 0 & \mathbb{D}_2 & \ddots & \\
\hline & \ddots & \ddots & 0 \\ \hline & & 0 & \mathbb{D}_n
\end{array}\right)\qquad{\rm with}\ \;\mathbb{D}_i\in SU(2)\,.$$

We can diagonalize each $SU(2)$ block $\mathbb{D}_i$ by
conjugating by some $\mathbb{P}_i\in SU(2)$. If we form the matrix
$P$ by placing these $SU(2)$ blocks, in order, down the diagonal
then we have $$P^{\dag}DP=X\qquad{\rm where}\ X\ {\rm is\
diagonal}\,.$$ Since $SU(2)=S\!p(2)$ we find that $P\in S\!p(2n)$.
Thus, for $A\in S\!p(2n)$, we have obtained the conjugation
$$Q^{\dag}P^{\dag}APQ=X\qquad{\rm where}\ Q,P,X\in S\!p(2n)\ {\rm
and}\ X\ {\rm is\ diagonal}\,.$$ Taking $U=PQ$, we have shown the
following to be true:

For all $A\in S\!p(2n)\ \exists\ U\in S\!p(2n)$ s.t. $U^{-1}AU=X$ where
$X$ is diagonal.

Now, recalling the set $\{E|E^{\dag}=E, E^2=I, JE^tJ=-E,
Tr(E)=2M-N\}$, we have $$E^2=I\quad {\rm and}\quad
JE^tJ=-E\Rightarrow EJE^t=J\,.$$ Thus $E\in S\!p(N)$ and we apply
our result above to give $E=UXU^{-1}$ for some $U\in S\!p(N)$. We
find that $X^2=I$ and $Tr(X)=2M-N$, so we have
$$X=\left(\begin{array}{cc} I_M & 0
\\ 0 & -I_{N-M}
\end{array}\right)\,.$$
(Note: any $X$ satisfying the above is conjugate to this $X$ via
some $S\!p(2n)$ matrix, which we absorb into $U$.) Thus we have
$E=UXU^{-1}$ with the required $X$.

\subsubsection{$\{U^{\ast}U^{\dag}|U\in SU(N)\}=\{E|E^{\dag}E=I, E^t=E,
\det{E}=1\}$}
\label{ssubsec:suso}

$\{U^{\ast}U^{\dag}|U\in SU(N)\}\subseteq\{E|E^{\dag}E=I, E^t=E, \det{E}=1\}$
\hspace{1em} is obvious.

Now if $E^{\dag}E=I$ then $\exists\ Q\in SU(N)$ s.t. $E=QDQ^{\dag}$,
where $D$ is diagonal.

If we impose the condition $E^t=E$, we have
$$Q^{\ast}DQ^t=QDQ^{\dag}\Rightarrow DQ^tQ=Q^tQD\,.$$ Since $D$ is
diagonal, we can find a diagonal matrix $C$ s.t. $C^2=D$ and with
the property $CQ^tQ=Q^tQC$. Thus, we have \beaa E&=&QC^2Q^{\dag}
\\ &=&Q^{\ast}Q^tQC^2Q^{\dag} \\ &=&Q^{\ast}CQ^tQCQ^{\dag} \\
&=&U^{\ast}U^{\dag}\qquad {\rm setting}\quad
Q^{\ast}CQ^t=U^{\ast}\,. \eeaa We see that
$\det{U}=\det{C^{\ast}}=\pm 1$, and so we have $U\in \{V\in
U(N)|\det{V}=\pm 1\}$.

Next, we show that the two sets $\{U^{\ast}U^{\dag}|UU^{\dag}=I,
\det{U}=1\}$ and $\{U^{\ast}U^{\dag}|UU^{\dag}=I, \det{U}=-1\}$
are in fact the same. This is trivially true for $N=1$, so we
assume $N\geq 2$ and suppose we have an element $E$ belonging to
the first set, that is $E=U^{\ast}U^{\dag}$ for $U\in SU(N)$. We
consider $U'=UT$ where $$T=\left(\begin{array}{cc|c} 0 & 1 & 0 \\
1 & 0 & \\ \hline \multicolumn{2}{c|}{0} & I_{N-2}
\end{array}\right)\qquad({\rm note}\colon\ T\in O(N),\ \det{T}=-1),$$
and we see that \beaa
{U'}^{\ast}{U'}^{\dag}&=U^{\ast}TT^tU^{\dag}\phantom{E} \\
&=U^{\ast}U^{\dag}\phantom{TT^tE} \\ &=E.\phantom{UTT^tU^t} \eeaa
We note that ${U'}^{\ast}{U'}^{\dag}$ is a member of the second
set, right multiplying by $T$ in this way is a self-inverse
operation, and thus we have shown the two sets to be equal.

Thus, we have established the equality $$\{U^{\ast}U^{\dag}|U\in
SU(N)\}=\{E|E^{\dag}E=I, E^t=E, \det{E}=1\}\,,$$
as required for section~\ref{ssubsec:cesuc}.

\subsubsection{$\{U^{\ast}JU^{\dag}|U\in U(N), \det{U}=\pm
1\}=\{E|E^{\dag}E=I, E^t=-E, \det{E}=1\}$}
\label{ssubsec:susp}

$\{U^{\ast}JU^{\dag}|U\in U(N), \det{U}=\pm 1\}\subseteq\{E|E^{\dag}E=I,
E^t=-E, \det{E}=1\}$\hspace{1em} is obvious.

Now if $E^{\dag}E=I$ then $\exists\ Q\in SU(N)$ s.t. $E=QDQ^{\dag}$,
where $D$ is diagonal.

If we impose the condition $E^t=-E$, we have
$$Q^{\ast}DQ^t=-QDQ^{\dag}\Rightarrow DQ^tQ=-Q^tQD\,.$$ We denote
the diagonal entries of $D$ by $D_i$, so that $D$ has entries
$D_i\delta_{ij}$. If we denote the entries of $Q^tQ$ by $P_{ij}$
then the condition above becomes
$$D_i\delta_{ij}P_{jk}=-P_{ij}D_j\delta_{jk}\Rightarrow
D_iP_{ik}=-D_kP_{ik}\,.$$ We see that $P_{ik}\neq 0\Rightarrow
D_i=-D_k$. By a suitable rearrangement of the diagonal entries of
$D$ (which we absorb by redefining $Q$) we take $D$ to be of the
form $$D=\left(\begin{array}{c|c|c|c|c} d_1I_{n_1} & 0 & \dots &
\dots & 0 \\ \hline 0 & -d_1I_{m_1} & 0 & & \vdots \\ \hline
\vdots & 0 & d_2I_{n_2} & \ddots & \vdots \\ \hline \vdots & &
\ddots & \ddots & 0 \\ \hline 0 & \dots & \dots & 0 & -d_kI_{m_k}
\end{array}\right)\quad\begin{array}{ll}
{\rm where} & n_i\geq m_i\geq 0, \\ & n_i\geq 1, \\ & d_j\neq\pm
d_i\ {\rm if}\ j\neq i.
\end{array}$$
Then $Q^tQ$ (which is symmetric) must have the form
$$Q^tQ=\left(\begin{array}{c|c|c|c|c|c|c}
0 & P_1 & 0 & 0 & \dots & 0 & 0 \\ \hline
P_1^t & 0 & 0 & 0 & \dots & 0 & 0 \\ \hline
0 & 0 & 0 & P_2 & & \vdots & \vdots \\ \hline
0 & 0 & P_2^t & 0 & & \vdots & \vdots \\ \hline
\vdots & \vdots & & & \ddots & \vdots & 0 \\ \hline
0 & 0 & \dots & \dots & \dots & 0 & P_k \\ \hline
0 & 0 & \dots & \dots & 0 & P_k^t & 0
\end{array}\right)\quad\begin{array}{ll}
{\rm where}\ P_i\ {\rm is\ of\ size} \\
n_i\ {\rm rows\ by}\ m_i\ {\rm columns.}
\end{array}$$
We recall that $Q^tQ\in SU(N)$, so that the $n_i$ rows of $P_i$
are orthonormal with respect to the inner product on
$\mathbb{C}^{m_i}$. Consequently, $m_i\geq n_i$ and so we must
have $n_i=m_i$, with $P_i\in SU(n_i)$, for all $i$.

We now decompose $D$ into two diagonal matrices $D=\tilde{D}\mathcal{E}$
where
$$\tilde{D}=\left(\begin{array}{c|c|c|c}
-id_1I_{2n_1} & 0 & \dots & 0\\ \hline
0 & -id_2I_{2n_2} & & \vdots \\ \hline
\vdots & & \ddots & 0 \\ \hline
0 & \dots & 0 & -id_kI_{2n_k}
\end{array}\right),\quad \mathcal{E}=\left(\begin{array}{cc|c|cc}
iI_{n_1} & 0 & & & \\
0 & -iI_{n_1} & & & \\ \hline
& & \ddots & & \\ \hline
& & & iI_{n_k} & 0 \\
& & & 0 & -iI_{n_k}
\end{array}\right).$$
These matrices satisfy
$$\tilde{D}Q^tQ=Q^tQ\tilde{D}\qquad\mathcal{E}Q^tQ=-Q^tQ\mathcal{E}\,.$$
Recall $\det{E}=1\Rightarrow\det{D}=1$, so since
$\det{\mathcal{E}}=1$ we also have $\det{\tilde{D}}=1$. We now
choose a diagonal matrix $C$ such that $C^2=\tilde{D}$ and
$CQ^tQ=Q^tQC$ whose determinant will be $\pm 1$. (Note that all
possible $C$ will have the same determinant.) Now we consider
\beaa E&=QDQ^{\dag}\phantom{Q^tQ^tMMMMMMMMMMMMMMMM} \\
&=QC^2\mathcal{E}Q^{\dag}\phantom{Q^tQ^tMMMMMMMMMMNNNNNN} \\
&=Q^{\ast}Q^tQC\mathcal{E}CQ^{\dag}\quad
C\mathcal{E}=\mathcal{E}C\ {\rm as}\ C\ {\rm and}\ \mathcal{E}\
{\rm are\ diagonal} \\
&=Q^{\ast}CQ^tQ\mathcal{E}CQ^{\dag}\,.\phantom{MMMMMMMMMMMMMMM}
\eeaa We set $R=Q\mathcal{E}Q^{\dag}$, then $Q\mathcal{E}=RQ$ so
that $E=Q^{\ast}CQ^tRQCQ^{\dag}$. Now we consider the properties
of $R$
$$R^{\dag}=Q\mathcal{E}^{\dag}Q^{\dag}=-Q\mathcal{E}Q^{\dag}=-R$$
$$R^t=Q^{\ast}\mathcal{E}Q^t=Q^{\ast}\mathcal{E}Q^tQQ^{\dag}=
-Q^{\ast}Q^tQ\mathcal{E}Q^{\dag}=-R\quad\Rightarrow R^{\ast}=R$$
$$R^2=Q\mathcal{E}Q^{\dag}Q\mathcal{E}Q^{\dag}=-I\quad{\rm as}\
\mathcal{E}^2=-I\,.$$ Further, $\det{R}=1$ so we have $R\in SO(N)$
and $R^2=-I$. Thus, appealing to the argument contained in the
next section~\ref{ssubsec:sou}, we can find a matrix $O\in O(N)$
such that $R=OJO^t$. Substituting this into our expression for $E$
we have $E=Q^{\ast}CQ^tOJO^tQCQ^{\dag}$. Setting
$U^{\ast}=Q^{\ast}CQ^tO\rightarrow U^{\dag}=O^tQCQ^{\dag}$ we have
$$E=U^{\ast}JU^{\dag}\qquad{\rm where}\ U\in U(N), \det{U}=\pm
1\,.$$

We have shown $\{U^{\ast}JU^{\dag}|U\in U(N), \det{U}=\pm
1\}=\{E|E^{ \dag}E=I, E^t=-E, \det{E}=1\}$, as required. However,
unlike the last subsection~\ref{ssubsec:suso}, here the
$\det{U}=\pm 1$ subsets are different, for consider: \beaa {\rm
suppose}\;\; U^{\ast}JU^{\dag}=V^{\ast}JV^{\dag}\ {\rm where}\
U,V\in U(N), \det{U}=1, \det{V}=-1\Rightarrow
\det{(V^tU^{\ast})}=-1:
\\{\rm then}\;\;
 U^{\ast}JU^{\dag}=V^{\ast}JV^{\dag}\Rightarrow
V^tU^{\ast}JU^{\dag}V=J \Rightarrow V^tU^{\ast}\in
S\!p(N)\Rightarrow \det{(V^tU^{\ast})}=1\quad \ctradn\phantom{MMM}
\eeaa

The subsets of $U(N)$ such that $\det=\pm 1$ are isomorphic via
multiplication by $\omega$ where $\omega^N=-1$. Thus
$$\{U^{\ast}JU^{\dag}|U\in U(N),
\det{U}=-1\}=\omega^2\{U^{\ast}JU^{ \dag}|U\in SU(N)\}$$ So we
have $$\{E|E^{\dag}E=I, E^t=-E,
\det{E}=1\}=\{1,\omega^2\}\times\{U^{\ast}J U^{\dag}|U\in
SU(N)\},$$
as required by subsection~\ref{ssubsec:cesuc}.

\subsubsection{$\{iUJU^t|U\in O(2n)\}=\{E|E^{\dag}=E, E^2=I, E^t=-E,
\}$}
\label{ssubsec:sou}

$\{iUJU^t|U\in O(2n)\}\subseteq\{E|E^{\dag}=E, E^2=I, E^t=-E,\}$\hspace{1em}
is obvious.

Now, $E^{\dag}=E, E^t=-E\Rightarrow E^{\ast}=-E$, so we consider
$F=-iE$. Then $F$ is a real matrix satisfying $$F^tF=I\qquad{\rm
and}\qquad F^2=-I\,.$$ Since $F$ is orthogonal there exists a
matrix $R\in SO(2n)$ such that $R^tFR$ has the form
$$\left(\begin{array}{c|c|c|c} O_1 & 0 & & \\ \hline 0 & O_2 &
\ddots & \\ \hline & \ddots & \ddots & 0 \\ \hline & & 0 & O_n
\end{array}\right)\qquad{\rm where}\ O_i\in O(2)\,.$$
Since $F^2=-I$ each $O_i^2=-I_2$, the only solutions for which $O_i\in
O(2)$ are
$$O_i=\pm \epsilon=\pm \left(\begin{array}{cc}
0 & -1 \\
1 & 0
\end{array}\right)\,.$$
We note that it is possible to conjugate $-\epsilon$ by the matrix
$$\left(\begin{array}{cc}
0 & 1 \\
1 & 0
\end{array}\right)\in O(2)$$
to obtain $\epsilon$. Thus we can find a matrix $R'\in
O(2)^{\otimes n}\subset O(2n)$ such that $R'^tR^tFRR'=J$. If we
set $U=RR'$ then we have $$E=iUJU^t\quad{\rm for}\ U\in
O(2n)\qquad{\rm as\ required.}$$

We note that $\{iUJU^t|U\in O(2n)\}\neq\{iUJU^t|U\in SO(2n)\}$,
since \beaa iUJU^t=iVJV^t\quad{\rm for}\ U,V\in O(N) \\
\Rightarrow V^tUJU^tV=J \phantom{MMMMiMMMM} \\ \Rightarrow V^tU\in
S\!p(2n) \phantom{MMMiMiMMMM} \\ \Rightarrow \det{U}=\det{V}\,.
\phantom{MMMiMiMMMM} \eeaa Both sets $\{iUJU^t|U\in SO(2n)\}$ and
$\{iUJU^t|U\in O(2n),\det{U}=-1\}$ are constructions of the
symmetric space $$\frac{SO(2n)}{U(n)}$$ and so $\{E|E^{\dag}=E,
E^2=I, E^t=-E\}$ is isomorphic to two copies of the symmetric
space, as stated in section~\ref{ssubsec:ceso}.

\subsubsection{$\{iUJU^{-1}|U\in S\!p(2n)\}=\{E|E^{\dag}=E, E^2=I,
JE^tJ=E\}$}
\label{ssubsec:spu}

$\{iUJU^{-1}|U\in S\!p(2n)\}\subseteq\{E|E^{\dag}=E, E^2=I, JE^tJ=E\}$
\hspace{1em} is obvious.

To establish the converse, we consider the conditions on $E$ \beaa
E^2=I\ {\rm and}\ JE^tJ=E\Rightarrow EJE^t=-J \\ E^{\dag}=E\ {\rm
and}\ E^2=I\Rightarrow E^{\dag}E=I\,. \eeaa If we let $F=-iE$ then
the conditions on $F$ are $$F^{\dag}F=I,\quad F^2=-I\quad{\rm
and}\quad FJF^t=J\,.$$ Thus $F\in S\!p(2n)$, so
from~\ref{ssubsec:spg} we can find a $V\in S\!p(2n)$ such that
$F=VDV^{-1}$, where $D$ is a diagonal matrix. Since $F^2=-I$ we
must also have $D^2=-I$, so the entries of $D$ must be $\pm i$. As
$D\in S\!p(2n)$ we cannot choose the signs of these diagonal
entries completely arbitrarily and we find we are restricted to
$$D=\left(\begin{array}{c|c|c|c} \pm \mathbb{I} & 0 & & \\ \hline
0 & \pm \mathbb{I} & \ddots & \\ \hline & \ddots & \ddots & 0 \\
\hline & & 0 & \pm \mathbb{I}
\end{array}\right)\,,\qquad{\rm recalling}\
\mathbb{I}=\left(\begin{array}{cc}
i & 0 \\
0 & -i
\end{array}\right)$$
and with each $\pm$ freely chosen. However, we have
$\mathbb{K}\,\mathbb{I}\mathbb{K}^{\dag}=-\mathbb{I}$ and so we
can further conjugate in $S\!p(2n)$ (which we absorb into $V$) to
ensure that $D$ has all $n\ \pm$ signs set to $+$.

Now we notice that $J\in S\!p(2n)$ and so, as above, there exists some
$W\in S\!p(2n)$ such that $D=WJW^{-1}$. If we set $U=VW$ then we have
obtained $F=UJU^{-1}$ where $U\in S\!p(2n)$. Recalling that $E=iF$, we
see that we have obtained
$$E=iUJU^{-1}\qquad{\rm as\ required.}$$

\section{Appendix: discrete ambiguities in boundary parameters}

First, recall the $G\!\times G$ invariance of the Lagrangian under
$g\mapsto g_L g g_R^{-1}$. Of course this should really be
${G\!\times G\over {\bf Z}(G)}$ (where ${\bf Z}(G)$ is the centre
of $G$, which is finite), since for $z\in {\bf Z}(G)$ we have
$g=zgz^{-1}$. The physics literature for the bulk principal chiral
model generally is not concerned with this. In the same way we do
not explore in the text the ambiguities in our boundary
parameters, but we wish here at least to state them precisely, and
to prove that they are finite in number.

\subsection{Chiral BCs}
\label{subsec:acbcs}

The ambiguity here arises because there may be non-trivial
$g_1,g_2$ such that $g_1 H g_2^{-1}=H$. This requires $$
\alpha(g_1) h \alpha(g_2^{-1}) = g_1hg_2^{-1} \qquad\forall h\in
H\,,$$ and thence $$ g_1^{-1}\alpha(g_1)
h=hg_2^{-1}\alpha(g_2)\qquad\forall h\in H.$$ So (by setting
$h=e$) we see that $$g_1^{-1}\alpha(g_1)=g_2^{-1}\alpha(g_2) =
k\qquad {\rm where} \qquad k\in {G\over H}\cap C(H)\,,$$ where
$C(H)<G$ is the centralizer of $H$. But, for a symmetric space,
$H$ is a maximal Lie subgroup (there is no Lie subgroup of greater
dimension which contains $H$), so ${G\over H}\cap C(H)$ is finite,
and so the solutions $g_1\in Hx,\,g_2\in Hy$ have $x=y$ (since the
Cartan immersion is 1-1) and are also finite in number.

\subsection{Non-chiral BCs}
\label{subsec:ancbcs}

Here the potential ambiguity is that there may be non-trivial
$g_1,g_2$ such that $g_1 {G\over H} g_2^{-1}={G\over H}$. In
contrast to the chiral case, we can push $g_1$ through ${G\over
H}$: \beaa g_1\{\alpha(g)g^{-1}|g\in G\} g_2^{-1} & = &
\{\alpha(\alpha(g_1)g)g^{-1}|g\in G\} g_2^{-1} \\ & = &
\{\alpha(\alpha(g_1)g)(\alpha(g_1)g)^{-1}|g\in G\}\alpha(g_1)
g_2^{-1} \\ & = & {G\over H}\alpha(g_1) g_2^{-1}\,. \eeaa So the
boundary is parametrized by $G$, up to $g_0$ such that ${G\over
H}g_0={G\over H}$. This requires $\alpha(kg_0) = g_0^{-1}k^{-1}$
for all $k\in G/H\hookrightarrow G$, so
$k^{-1}\alpha(g_0)=g_0^{-1} k^{-1}$. This must hold for $k=e$, so
$g_0\in {\cal M}$ of (\ref{EFman}), and commutes with every
element of $G/H\hookrightarrow G$.

Such $g_0$ form a group, which must be finite: for suppose not,
that its algebra is generated by $\kk_0\subset\kk$. Then
$[\kk_0,\kk]=0$. Also $[[\h,\kk_0],\kk]\subset
[[\kk,\kk_0],\h]+[[\h,\kk],\kk_0]$, both of which are empty, so
$[\h,\kk_0]\subset\kk_0$. Thus $[\kk_0,\g]\subset\kk_0$, and
$\kk_0$ is an ideal, and is therefore trivial.

Note the specialization (mentioned in the text) when $g_1=g_2$:
the boundary is then parametrized by $\alpha(g_1)g_1^{-1}$ and
thus by $G/H$ (again quotiented, here by ${\bf Z}(G/H)$, those
elements of $G/H$ which commute with all of $G/H$). It is
straightforward to propose a compatible PCM boundary $S$-matrix,
though we do not do so here.

\section{Appendix: boundary Yang Baxter and crossing-unitarity
calculations}
\label{sec:bybcuc}

In this section we include a representative selection of the BYBE
and crossing-unitarity calculations required to obtain the various
constraints on the boundary $S$-matrices that we have considered
in this paper. We hope they will be sufficiently illustrative that
the interested reader can perform any calculations not presented
here for themselves.

First we list the minimal bulk $S$-matrices derived
in~\cite{ogiev87} \beaa SU(N) : &
{\displaystyle\frac{\sigma_u(\theta)}{(1-\frac{h\theta}{2i\pi})}}
\left(\tws-{\displaystyle\frac{h\theta}{2i\pi}}\twc\right) \\
SO(N) : &
{\displaystyle\frac{\sigma_o(\theta)}{(1-\frac{h\theta}{2i\pi})}}
\left(\tws-{\displaystyle\frac{h\theta}{2i\pi}}\twc+{\displaystyle\frac{
\frac{h\theta}{2i\pi}}{(\frac{h}{2}-\frac{h\theta}{2i\pi})}}\twu\right)
\\ S\!p(N) : &
{\displaystyle\frac{\sigma_p(\theta)}{(1-\frac{h\theta}{2i\pi})}}
\left(\tws-{\displaystyle\frac{h\theta}{2i\pi}}\twc+{\displaystyle\frac{
\frac{h\theta}{2i\pi}}{(\frac{h}{2}-\frac{h\theta}{2i\pi})}}\twj\right)
\eeaa where $h$ is the dual Coxeter number and the scalar
prefactors $\sigma$ are~\footnote{The $\mp$ sign in the prefactor for
$SU(N)$ reflects the fact that this sign is an arbitrary choice.
However, from consideration of the boundary bootstrap, details of which
are outside the scope of this paper, the indications are that a $-$ sign
is required for the model with non-conjugating boundary conditions,
whilst a $+$ sign is the preferred choice for representation conjugating
BCs.}
\beaa
\sigma_u(\theta)=\mp\frac{\Gamma\left(\frac{\theta}{2i\pi}+
\frac{1}{h}\right)\Gamma\left(\frac{-\theta}{2i\pi}\right)}{\Gamma
\left(\frac{-\theta}{2i\pi}+\frac{1}{h} \right)\Gamma
\left(\frac{\theta }{2i\pi}\right)}\phantom{MMMMMMM} \\
\sigma_o(\theta)=-\frac{\Gamma\left(\frac{
\theta}{2i\pi}+\frac{1}{h}\right)\Gamma\left(\frac{\theta}{2i\pi}+\frac{
1}{2}\right)\Gamma\left(\frac{-\theta}{2i\pi}\right)\Gamma\left(\frac{
-\theta}{2i\pi}+\frac{1}{2}+\frac{1}{h}\right)}{\Gamma\left(\frac{
-\theta}{2i\pi}+\frac{1}{h}\right)\Gamma\left(\frac{-\theta}{2i\pi}+
\frac{1}{2}\right)\Gamma\left(\frac{\theta}{2i\pi}\right)\Gamma\left(
\frac{\theta}{2i\pi}+\frac{1}{2}+\frac{1}{h}\right)} \\
\sigma_p(\theta)=-\frac{\Gamma\left(\frac{
\theta}{2i\pi}+\frac{1}{h}\right)\Gamma\left(\frac{\theta}{2i\pi}+\frac{
1}{2}\right)\Gamma\left(\frac{-\theta}{2i\pi}\right)\Gamma\left(\frac{
-\theta}{2i\pi}+\frac{1}{2}+\frac{1}{h}\right)}{\Gamma\left(\frac{
-\theta}{2i\pi}+\frac{1}{h}\right)\Gamma\left(\frac{-\theta}{2i\pi}+
\frac{1}{2}\right)\Gamma\left(\frac{\theta}{2i\pi}\right)\Gamma\left(
\frac{\theta}{2i\pi}+\frac{1}{2}+\frac{1}{h}\right)} \eeaa where
$\Gamma$ is the gamma function, and $$h=\left\{\begin{array}{ll} N
& SU(N) \\ N-2 & SO(N) \\ N+2 & S\!p(N)\,.
\end{array}\right.
$$ Note that $$\bar{\sigma}_u(\theta)=\sigma_u(i\pi-\theta)\,.$$

\subsection{BYBE calculations}
\label{subsec:bybecal}

Recall the boundary Yang Baxter equation \beaa
S_{ij}^{kl}(\theta-\phi)\left(I_{jm}\otimes K^{ln}(\theta)\right)
S_{mo}^{np}(\theta+\phi)\left(I_{oq}\otimes K^{pr}(\phi)\right)=
\\ \left(I_{ij}\otimes K^{kl}(\phi)\right)S_{jm}^{ln}(\theta+\phi)
\left(I_{mo}\otimes
K^{np}(\theta)\right)S_{oq}^{pr}(\theta-\phi)\,. \eeaa For clarity
of the calculations we introduce the notation
$$u=\frac{h\theta}{2i\pi},\qquad v=\frac{h\phi}{2i\pi},\qquad
u_0=\frac{h}{4}.$$ In any BYBE calculation the scalar prefactors
cancel, and we consider here only the matrix part of the equation.

\subsubsection{The $SU(N)$ case with $K_1$ boundary $S$-matrix}

Substituting into the BYBE with the bulk $S$-matrix for the
$SU(N)$ PCM and $K_1$ gives \beaa \left( \tws -(u-v) \twc \right)
\twssu \left( \tws -(u+v) \twc \right) \twssu = \\ \twssu \left(
\tws -(u+v) \twc \right) \twssu \left( \tws -(u-v) \twc \right)
\eeaa Expanding out and cancelling where possible, we are left
with $$\twc \twssu \twssu = \twssu \twssu \twc$$ Thus, for the
equation to be satisfied we require the condition~:- $$\ons \ons
=\alpha \one \quad {\rm for\ some\ constant}\ \alpha .$$

\subsubsection{The $S\!p(N)$ case with $K_2$ boundary $S$-matrix}

Substituting into the BYBE with the bulk $S$-matrix for the
$S\!p(N)$ PCM and $K_2$ gives \beaa
\left(\tws-(u-v)\twc+t(u-v)\twj\right)\left(\tws+\tilde{c}u\twssu\right)\left(
\tws-(u+v)\twc+t(u+v)\twj\right)\times \\
\left(\tws+\tilde{c}v\twssu\right)\ =\
\left(\tws+\tilde{c}v\twssu\right)\times \phantom{MMMMMMMMM} \\
\left(\tws-(u+v)\twc+t(u+v)\twj\right)\left(\tws+\tilde{c}u\twssu\right)
\left(\tws-(u-v)\twc+t(u-v)\twj\right) \eeaa where
$t(u)=\frac{u}{2u_0-u}$ and $\tilde{c}$ is related to the original
$S$-matrix constant $c$ by the relation $\tilde{c}=\frac{2i\pi
c}{h}$. Expanding out, cancelling where possible (noting that the
terms involving $\twssu\twc$ and $\twc\twssu$ cancel after some
simple algebra) and rearranging (some less trivial algebra!) we
are left with \beaa -\tilde{c}uv
t(u-v)t(u+v)(2+\tilde{c}\trs)\left(\twj\twssu-\twssu\twj\right) \\
+2\tilde{c}uvt(u-v)t(u+v)\left(\twj\twssu\twc-\twc\twssu\twj\right)
\\
\tilde{c}^2uv(u-v)\left(\twssu\twssu\twc-\twc\twssu\twssu\right)+
\tilde{c}^2uvt(u-v) \left(\twj\twssu\twssu-\twssu\twssu\twj\right)
\\
-\tilde{c}^2uv(u+v)t(u-v)\left(\twj\twssu\twc\twssu-\twssu\twc\twssu
\twj\right) \\
-\tilde{c}^2uv(u-v)t(u+v)\left(\twc\twssu\twj\twssu-\twssu\twj\twssu
\twc\right) =0\,. \eeaa In order for this to hold we are forced to
have \beaa \ons \ons &=\alpha \one \quad {\rm for\ some\
constant}\ \alpha \\ \twc \twssu \onjdl &=\beta \twssu \onjdl
\quad {\rm i.e.}\ \quad ( \ons \onjr )^t =\beta \ons \onjr \eeaa
(Note: then ${\beta}^2=1$.) We find that the equation is then
satisfied provided $$2 \beta -2-\tilde{c}\trs=0\,.$$ Since $\beta
= \pm 1$ we must have~:- $$\ons\ons=\alpha\one\qquad{\rm and}$$
$$(\ons\onjr)^t=\ons\onjr\Longleftrightarrow
\tilde{c}\trs=0\quad{\rm or}\quad
(\ons\onjr)^t=-(\ons\onjr)\Longleftrightarrow
\tilde{c}\trs=-4\,.$$

\subsubsection{The $SU(N)$-conjugating case}

Recall the conjugated BYBE \beaa S_{ij}^{kl}(\theta-\phi)( I_{jm}
\otimes K^{l\bar{n}}(\theta))
S_{m\bar{o}}^{\bar{n}p}(\theta+\phi)( I_{\bar{o}\bar{q}} \otimes
K^{p\bar{r}}(\phi))= \phantom{MMMMMMMM} \\ \phantom{MMMMMMMM}
(I_{ij} \otimes
K^{k\bar{l}}(\phi))S_{j\bar{m}}^{\bar{l}n}(\theta+\phi)
(I_{\bar{m}\bar{o}} \otimes K^{n\bar{p}}(\theta))
S_{\bar{o}\bar{q}}^{\bar{p}\bar{r}}(\theta-\phi)\,. \eeaa The
$K_2$ boundary $S$-matrix does not satisfy the above equation, but
$K_1$ does, under some constraints. Substituting in and using our
simplifying notation we get \beaa
\left(\tws-(u-v)\twc\right)\twssu\left(\twu-(2u_0-u-v)\twc\right)\twssu=
\\
\twssu\left(\twu-(2u_0-u-v)\twc\right)\twssu\left(\tws-(u-v)\twc\right)
\eeaa
Expanding out and cancelling all possible terms leaves
$$\twc\twssu\twu\twssu=\twssu\twu\twssu\twc$$
So to satisfy the conjugated BYBE we must have $(\ons)^t=\pm\ons$.

There are two other BYBEs to consider in addition to the $V\otimes
V\to\bar{V}\otimes\bar{V}$ case considered so far. The
$\bar{V}\otimes\bar{V}\to V\otimes V$ BYBE will be similar to the
above, so that if we denote by $\onsc$ the matrix part of the
$\bar{V}\to V$ boundary $S$-matrix then we must have
$(\onsc)^t=\pm\onsc$. The last case to consider is
$V\otimes\bar{V}\to\bar{V}\otimes V$, where the BYBE is \beaa
S_{\bar{i}j}^{k\bar{l}}(\theta-\phi)\left(I_{jm}\otimes
K^{\bar{l}n}(
\theta)\right)S_{mo}^{np}(\theta+\phi)\left(I_{oq}\otimes
K^{p\bar{r}}( \phi)\right)= \phantom{MMMMMMMM} \\
\phantom{MMMMMMMM} \left(I_{\bar{i}\bar{j}}\otimes
K^{k\bar{l}}(\phi)\right)S_{\bar{j}\bar{
m}}^{\bar{l}\bar{n}}(\theta+\phi)\left(I_{\bar{m}\bar{o}}\otimes
K^{\bar{n}p}(\theta)\right)S_{\bar{o}q}^{p\bar{r}}(\theta-\phi)\,.
\eeaa Substituting into this, again with simplified notation, we
get \beaa \left( \twu -(2u_0-u+v) \twc \right) \left( \twsscu
\right) \left( \tws -(u+v) \twc \right) \left( \twssu \right)= \\
\left( \twssu \right) \left( \tws -(u+v) \twc \right) \left(
\twsscu \right) \left( \twu -(2u_0-u+v) \twc \right)\,. \eeaa
Expanding out and cancelling all the terms we can we have \beaa
\twu\twsscu\twssu-(2u_0-u+v)\twc\twsscu\twssu-(u+v)\twu\twsscu\twc
\twssu= \\
\twssu\twsscu\twu-(2u_0-u+v)\twssu\twsscu\twc-(u+v)\twssu\twc\twsscu\twu\,.
\eeaa This equation is satisfied provided $$\ons \onsc = \onsc
\ons = \alpha \one \quad {\rm and} \quad \twssu \twc \twsscu \onul
= \twsscu \twc \twssu \onul = \beta \onul\,.$$ Since $(\ons)^t=\pm
\ons$ and $(\onsc)^t=\pm \onsc$, we have $\beta=\pm \alpha$ and so
have only one independent parameter. The conditions imposed for
the conjugated $SU(N)$ case are thus $$(\ons)^t=\pm\ons\quad {\rm
and}\quad(\onsc)^t=\pm\onsc\quad{\rm
and}\quad\ons\onsc=\onsc\ons=\alpha\one\,.$$

\subsection{Crossing-unitarity calculations}
\label{subsec:cucal}

Recall the crossing-unitarity equation

$$K^{ij}(\frac{i\pi}{2}-\theta)=S_{\bar{j}k}^{i\bar{l}}(2\theta)
K^{\bar{l}\bar{k}}(\frac{i\pi}{2}+\theta)\,.$$
We note that it is $S_{\bar{j}k}^{i\bar{l}}(2\theta)$ that is
required here, which is the crossed $S$-matrix. We obtain it by taking
the standard $S$-matrix substituting $i\pi-2\theta$ for $2\theta$ and
turning the matrix diagrams through $90^o$. (We note that the process
of crossing doesn't alter the $S$-matrix for the $SO(N)$ and $S\!p(N)$
cases, but we go through the process anyway to illustrate the $SU(N)$
cases.) We again make use of some simplifying notation.

\subsubsection{The $SO(N)$ case with $K_2$ boundary $S$-matrix}

Substituting into the crossing-unitarity equation we have \beaa
\frac{\tau(\frac{i\pi}{2}-\theta)}{(1-c(\frac{i\pi}{2}-\theta))}\left(
\onul+\tilde{c}(u_0-u)\onuls\right)=\frac{\sigma_o(i\pi-2\theta)\tau(\frac{
i\pi}{2}+\theta)}{(1-2u_o+2u)(1-c(\frac{i\pi}{2}+\theta))}\times
\phantom{MMMMMM} \\
\phantom{MMMMMM}\left(\twu-2(u_0-u)\twc+\frac{u_0-u}{u}\tws\right)\left(
\onul+\tilde{c}(u_0+u)\onuls\right) \eeaa \beaa
\Rightarrow\frac{\tau(\frac{i\pi}{2}-\theta)}{\tau(\frac{i\pi}{2}+
\theta)}\left(\onul+\tilde{c}(u_0-u)\onuls\right)=\frac{\begin{array}{c}
\sigma_o(i\pi-2\theta)\times \\ (1-c(\frac{i\pi}{2}-\theta))
\end{array}}{\begin{array}{c}
(1-2u_o+2u)\times \\
(1-c(\frac{i\pi}{2}+\theta))
\end{array}}\left(\begin{array}{c}
\left(\begin{array}{c} N+\tilde{c}\trs(u_0+u) \\
-2(u_0-u)+{\displaystyle\frac{u_0-u}{u}}
\end{array}\right)\onul \\
-2\tilde{c}(u_0-u)(u_0+u)\twc\onuls \\
+{\displaystyle\frac{\tilde{c}(u_0-u)(u_0+u)}{u}}\onuls
\end{array}\right)
\eeaa In order for this to be satisfied it is necessary to impose
$(\ons)^t=\pm\ons$. Considering coefficients of the $\onuls$ terms
we find
$$\frac{\tau(\frac{i\pi}{2}-\theta)}{\tau(\frac{i\pi}{2}+\theta)}=
\frac{\sigma_o(i\pi-2\theta)(u_0+u)}{(1-2u_0+2u)}\left(\frac{1}{u}\mp 2
\right)\frac{(1-c(\frac{i\pi}{2}-\theta))}{(1-c(\frac{i\pi}{2}+\theta))}$$
For the coefficients of the $\onul$ terms to be consistent with
this we require a constraint on $\trs$ which depends on the choice of $\pm$,
altogether we have the matrix constraints
$$(\ons)^t=\ons\Longleftrightarrow\tilde{c}\trs=-4\quad{\rm or}\quad(
\ons)^t=-(\ons)\Longleftrightarrow\tilde{c}\trs=0.$$
From the crossing symmetry of the $S$-matrix we can simplify the
constraint on the scalar prefactor, obtaining
$$\frac{\tau(\frac{i\pi}{2}-\theta)}{\tau(\frac{i\pi}{2}+\theta)}=
\frac{(u_0+u)(1\mp 2u)(1-c(\frac{i\pi}{2}-\theta))}{(u_0-u)(1-2u)(1
-c(\frac{i\pi}{2}+\theta))}\sigma_o(2\theta)$$ which can be written as
$$\frac{\tau(\frac{i\pi}{2}-\theta)}{\tau(\frac{i\pi}{2}+\theta)}=
\left\{\begin{array}{ll}
\left[\frac{h}{2}\right]\left[\frac{h}{ci\pi}-\frac{h}{2}\right]
\sigma_o(2\theta) & (\ons)^t=\ons \\
-[1]\left[\frac{h}{2}\right]\left[\frac{h}{ci\pi}-\frac{h}{2}\right]
\sigma_o(2\theta) & (\ons)^t=-\ons\,.
\end{array}\right.$$

\subsubsection{The $SU(N)$-conjugating case}

The $K$-matrix for $SU(N)$-conjugating must satisfy a conjugated
version of the crossing-unitarity equation,
$$K^{i\bar{j}}(\frac{i\pi}{2}-\theta)=S_{jk}^{il}(2\theta)K^{l\bar{k}}
(\frac{i\pi}{2}+\theta)\,.$$ For this case it is not
the crossed $S$-matrix we require, but the standard $S$-matrix.
Substituting in, we have
$$\rho(\frac{i\pi}{2}-\theta)\onuls=\frac{\sigma_u(2\theta)}{(1-2u)}
\left(\tws-2u\twc\right)\rho(\frac{i\pi}{2}+\theta)\onuls\,.$$ On
expanding this, we see that $(\ons)^t=\pm\ons$ is required. Then
the condition on the scalar prefactor becomes
$$\frac{\rho(\frac{i\pi}{2}-\theta)}{\rho(\frac{i\pi}{2}+\theta)}=
\frac{(1\mp 2u)}{(1-2u)}\sigma_u(2\theta)\,,$$ which can be
written as
$$\frac{\rho(\frac{i\pi}{2}-\theta)}{\rho(\frac{i\pi}{2}+\theta)}=
\left\{\begin{array}{ll} \sigma_u(2\theta) & (\ons)^t=\ons \\
-[1]\sigma_u(2\theta) & (\ons)^t=-\ons\,.
\end{array}\right.$$

\section{Appendix: unitarity and hermitian analyticity calculations}
\label{sec:apuhac}

In this section we prove the two results, concerning complex parameters
that could consistently be set to $1$, stated in
section~\ref{subsec:uha}.

\subsection{The non-conjugating case}
\label{subsec:hanc}

We start from~(\ref{mbsmun}), $$E^{\dag}=E\quad{\rm and}\quad
E^2=\alpha I\quad{\rm where}\ \; \alpha\in U(1).$$ Since $E$ is
hermitian we can diagonalize it as $$D=QEQ^{\dag}\;\;{\rm
where}\;\; Q\in SU(N).$$ Then we have
$$D^{\dag}=QE^{\dag}Q^{\dag}=QEQ^{\dag}=D\Rightarrow D^{\ast}=D.$$
Further, $$D^2=QEQ^{\dag}QEQ^{\dag}=QE^2Q^{\dag}=\alpha I.$$ Now
$D^{\ast}=D\Rightarrow \alpha\in\mathbb{R^+}$ so $\alpha=1$, as
stated in~\ref{ssubsec:uhanc}.

\subsection{The conjugating case}
\label{subsec:hac}

We start from~(\ref{mbsmuc1})
\beaa
\alpha E=F^{\dag}\phantom{AEAI}&{\rm and}&\ \quad\rho(\theta)=\alpha
\omega(-\theta^{\ast})^{\ast} \\
EF=\beta I\phantom{AEA^{\dag}}&{\rm and}&\ \quad\rho(\theta)\omega(
-\theta)=\frac{1}{\beta}\,. \eeaa
We can use the rescaling freedom in $K(\theta)=\rho(\theta)E$ and
$K'(\theta)=\omega(\theta)F$,
\beaa
E\to\lambda E, & \rho(\theta)\to\frac{1}{\lambda}\rho(\theta), \\
F\to\kappa F, & \omega(\theta)\to\frac{1}{\kappa}\omega(\theta),
\eeaa
to set both $\alpha$ and $\beta$ to $1$. Once this has been done some
rescaling freedom still remains: the phase shift $\lambda=e^{i\psi}$,
$\kappa=e^{-i\psi}$ leaves the constraints unchanged. Thus, we can also
insist that $\det{E}=1\Longleftrightarrow\det{F}=1$ as stated
in~\ref{ssubsec:uhac}.

\end{document}